# The Impact of Automated Parameter Optimization on Defect Prediction Models

Chakkrit Tantithamthavorn, *Member, IEEE,* Shane McIntosh, *Member, IEEE,* Ahmed E. Hassan, *Senior Member, IEEE,* and Kenichi Matsumoto, *Senior Member, IEEE*

**Abstract**— Defect prediction models—classifiers that identify defect-prone software modules—have configurable parameters that control their characteristics (e.g., the number of trees in a random forest). Recent studies show that these classifiers underperform when default settings are used. In this paper, we study the impact of automated parameter optimization on defect prediction models. Through a case study of 18 datasets, we find that automated parameter optimization: (1) improves AUC performance by up to 40 percentage points; (2) yields classifiers that are at least as stable as those trained using default settings; (3) substantially shifts the importance ranking of variables, with as few as 28% of the top-ranked variables in optimized classifiers also being top-ranked in non-optimized classifiers; (4) yields optimized settings for 17 of the 20 most sensitive parameters that transfer among datasets without a statistically significant drop in performance; and (5) adds less than 30 minutes of additional computation to 12 of the 26 studied classification techniques. While widely-used classification techniques like random forest and support vector machines are not optimization-sensitive, traditionally overlooked techniques like C5.0 and neural networks can actually outperform widely-used techniques after optimization is applied. This highlights the importance of exploring the parameter space when using parameter-sensitive classification techniques.

## 1 Introduction

The limited Software Quality Assurance (SQA) resources of software organizations must focus on software modules (e.g., source code files) that are likely to be defective in the future. To that end, defect prediction models are trained to identify defect-prone software modules using statistical or machine learning classification techniques.

Such classification techniques often have configurable parameters that control the characteristics of the classifiers that they produce. For example, the number of decision trees of which a random forest classifier is comprised can be configured prior to training the forest. Furthermore, the number of non-overlapping clusters of which a $k$-nearest neighbours classifier is comprised must be configured prior to using the classification technique.

Since the optimal settings for these parameters are not known ahead of time, the settings are often left at default values. Moreover, automated parameter optimization may increase the risk of *overfitting*, i.e., producing classifiers that are too specialized for the data from which they were trained. However, recent work suggests that defect prediction models may underperform if they are trained using suboptimal parameter settings. For example, Jiang *et al.* [63] and Tosun *et al.* [146] point out that the default parameter settings of random forest and naïve bayes are often suboptimal. Koru *et al.* [42] and Mende *et al.* [93, 94] show that selecting different parameter settings can impact the performance of defect models. Hall *et al.* [45] show that unstable classification techniques may underperform due to the use of default parameter settings. Mittas *et al.* [98] and Menzies *et al.* [97] argue that unstable classification techniques (e.g., classifiers that are sensitive to parameter settings) make replication of defect prediction studies more difficult.

Indeed, we perform a literature analysis that reveals that 26 of the 30 most commonly used classification techniques (87%) in software defect prediction studies require at least one parameter setting. Fu *et al.* [35] also point out that 80% of the 50 highest cited defect prediction studies rely on a default setting that is provided by the employed research toolkits (e.g., R or Weka). Since such parameter settings may impact the performance of defect prediction models, the settings should be carefully selected. However, it is impractical to assess all of the possible settings in the parameter space of a single classification technique [10, 46, 73]. For example, Kocaguneli *et al.* [73] show that there are at least 17,000 possible settings of the combinations of 6 parameters to explore when training a $k$-nearest neighbours classifier for effort estimation models (i.e., 3 case subset selectors, 8 feature weighting methods, 5 discretization methods, 6 similarity measures, 4 adoption mechanisms, and 6 analogies).

In this paper, we investigate the (1) performance improvements, (2) performance stability (i.e., the variance of the performance estimates that are derived from defect pre-

- C. Tantithamthavorn is with the School of Computer Science, the University of Adelaide, Australia.
E-mail: chakkrit.tantithamthavorn@adelaide.edu.au.
- S. McIntosh is with the Department of Electrical and Computer Engineering, McGill University, Canada. E-mail: shane.mcintosh@mcgill.ca.
- A. E. Hassan is with the School of Computing, Queen's University, Canada. E-mail: ahmed@cs.queensu.ca.
- K. Matsumoto is with the Graduate School of Information Science, Nara Institute of Science and Technology, Japan. E-mail: matumoto@is.naist.jp.

diction models), (3) model interpretation (i.e., the ranking of the importance of variables), (4) parameter transferability, (5) computational cost, and (6) and the ranking of classification techniques of defect prediction models where automated parameter optimization techniques (e.g., grid search) are applied. The four automated parameter optimization techniques evaluate candidate parameter settings and suggests the optimized settings that achieve the highest performance. We evaluate 26 commonly-used classification techniques using 12 performance measures, namely, 3 threshold-independent (e.g., AUC) and 9 threshold-dependent (e.g., precision, recall) performance measures. Furthermore, we also conduct an empirical comparison of the four automated parameter optimization techniques (i.e., grid search, random search, genetic algorithm, differential evolution) for defect prediction models. Through a case study of 18 datasets from systems that span both proprietary and open source domains, we record our observations with respect to seven dimensions:

**(1) Performance improvement:** Optimization improves the AUC performance of defect prediction models by up to 40 percentage points (e.g., C5.0 classifier). The performance improvement when applying optimization is non-negligible for 16 of the 26 studied classification techniques (62%), for example, C5.0, neural networks, and CART classifiers. Fortunately, the optimization of the parameters of random forest classifiers, one of the most popularly-used techniques in defect prediction studies, tends to have a negligible to small impact on the performance of random forest. Nevertheless, the impact of parameter optimization should be carefully considered when researchers select a particular technique for their studies.

**(2) Performance stability:** Optimized classifiers are at least as stable as classifiers that are trained using the default settings. Moreover, the optimized classifiers of 9 (including C5.0, neural networks, and CART classifiers) of the 26 studied classification techniques (35%) are significantly more stable than classifiers that are trained using the default settings. On the other hand, for random forest and naïve bayes classifiers, the performance of the default-settings models is as stable as the optimized models.

**(3) Model interpretation:** Optimization substantially shifts the importance ranking of variables, with as few as 28% of the top-ranked variables from the optimized models appearing in the same ranks as the non-optimized models. Classification techniques where optimization has a large impact on performance (e.g., C5.0, neural networks, CART classifiers) are subject to large shifts in interpretation. On the other hand, 95% of the variables appear at the same rank in the classification techniques where optimization has a negligible performance improvement (e.g., random forest classifiers).

**(4) Parameter transferability:** The optimal settings of 17 of the 20 most sensitive parameters can be applied to datasets that share a similar set of metrics without a statistically significant drop in performance. On the other hand, we find that only the parameters of LogitBoost, FDA, and xGBTree cannot be transferred across any of the studied datasets, indicating that researchers and practitioners should re-apply automated parameter optimization for such classifiers. However, we note that classifiers with low transferability are not top-performing classifiers.

**(5) Computational cost:** Grid-search optimization adds less than 30 minutes of additional computation time to 12 of the 26 studied classification techniques. To assign an actual monetary cost for such optimization techniques, we used an Amazon EC2 costing estimates. We find that an Amazon EC2's estimated computational cost of the 3 slowest classification techniques is less than $1.

**(6) The ranking of classification techniques:** Random forest is not the top-performing classification technique for all of the studied datasets, which disagrees with the findings of prior general data mining and recent software engineering related studies. In addition, some rarely-used low-performing classification techniques (e.g., C5.0) are substantially outperforming widely-used techniques like random forest—*highlighting the importance of exploring the parameter settings of classifiers.*

**(7) Empirical comparison of automated parameter optimization techniques:** Grid search, random search, genetic algorithm, and differential evolution techniques yield similar benefits of performance improvement, when applying to defect prediction models irrespective of the choice of performance measures.

Our results lead us to conclude that automated parameter optimization can have a large impact on the performance, stability, interpretation, and ranking of defect prediction models for many commonly-used classification techniques like neural networks. However, there are some classification techniques like random forest and support vector machines whose parameter optimization has a negligible to small improvement. Thus, future studies should apply automated parameter optimization only if the used classification techniques are sensitive to parameter settings, and should not be too concerned about the parameter optimization of one of most commonly-used classification techniques (i.e., random forest). Since we find that random forest tends to be the top-performing classification techniques, tends to produce stable performance estimates, and is insensitive to parameter settings, future studies should consider using a random forest classifier when constructing defect prediction models.

Finally, we would like to emphasize that we do not seek to claim the generalization of our results. Instead, the key message of our study is to shed light that automated parameter optimization should be applied in many classification techniques like neural networks, but it is not essential in some classification techniques like random forest. Moreover, some rarely-used low-performing classification techniques (e.g., C5.0) are substantially outperforming widely-used techniques like random forest—highlighting the importance of exploring the parameter settings of classifiers.

While recent studies throughout the community have used our prior work [139] to advocate the need for applying parameter optimization to all classifiers, the key message of our extend analyses, as presented in this paper, is to highlight that parameter optimization is not essential for all classifiers. For example, some very commonly-used clas-sifiers like random forest are not impacted by our prior findings. However, some rarely used classifiers are able to outperform commonly-used classifiers. Hence, future studies might want to consider the C5.0 classifier. Furthermore, the reported AUC performance and analyses of random forest of prior studies is likely not impacted by our recent work [139]. Nevertheless, more improved performance might be possible in such prior studies if classifiers such as C5.0, and neural networks were considered (and whenever if appropriate). This paper shows that C5.0 performs as well and is as stable as random forest.

### 1.1 Contributions

This paper is an extended version of our earlier work [139] (i.e., RQs 1, 2, and 6). This extension makes eight additional conceptual contributions and two technical contributions:

1) An extensive investigation of the impact of an automated parameter optimization on 12 performance measures, namely, 3 threshold-independent (e.g., AUC) and 9 threshold-dependent (e.g., precision, recall) performance measures (RQ1).
2) An investigation of the impact of parameter settings of classification techniques on the interpretation of defect prediction models when automated parameter optimization is applied (RQ3).
3) An investigation of parameter transferability across defect datasets when automated parameter optimization is applied (RQ4).
4) An investigation of the computational costs of applying automated parameter optimization (RQ5).
5) An in-depth discussion of the ranking of optimized and default-settings classifiers for defect prediction models (RQ6).
6) An introduction to the state-of-the-art automated parameter optimization techniques for defect prediction models (Section 8.1).
7) An empirical comparison of automated parameter optimization techniques for defect prediction models (Section 8.2.3).
8) An in-depth and a critical discussion of the broader implications of our findings with respect to (1) the conclusions of prior defect prediction studies; (2) the experimental design of defect prediction models; and (3) search-based software engineering literature (Section 9).

Our two technical contributions are as follows:

1) A generic variable importance calculation that applies to the 26 studied classification techniques (Section 5.5).
2) An improvement to the Scott-Knott ESD test (v2.0) (Section 5.5.3). The implementation of the improved ScottKnottESD test (v2.0) is also available on GitHub[1] and CRAN[2] [134].

### 1.2 Paper Organization

The remainder of the paper is organized as follows. Section 2 illustrates the relevance of the parameter settings of classification techniques for defect prediction models. Section 3 positions this paper with respect to the related work. Section 4 discusses our selection criteria of the studied systems. Section 5 presents the design of our case study, while Section 6 presents the results. Section 7 revisits prior analyses of the ranking of classification techniques when automated parameter optimization is applied. Section 8 presents the approach and the results of an empirical comparison of four automated parameter optimization techniques for defect prediction models. Section 9 discusses the broader implications of our findings. Section 10 discloses the threats to the validity of our study. Finally, Section 11 draws conclusions.

1. http://github.com/klainfo/ScottKnottESD
2. https://cran.r-project.org/web/packages/ScottKnottESD/

## 2 The Relevance of Parameter Settings for Defect prediction models

A variety of classification techniques are used to train defect prediction models. Since some classification techniques do not require parameter settings (e.g., logistic regression), we first assess whether the most commonly used classification techniques require parameter settings.

We begin with the six families of classification techniques that are used by Lessmann *et al.* [81]. Based on a recent literature review [80], we add five additional families of classification techniques that were recently used in defect prediction studies (e.g., Ghotra *et al.* [39, 40]). In total, we study 30 classification techniques that span 11 classifier families. Table 1 provides an overview of the 11 families of classification techniques.

Our literature analysis reveals that 26 of the 30 most commonly used classification techniques require at least one parameter setting. Table 1 provides an overview of the 25 unique parameters that apply to the studied classification techniques.

*26 of the 30 most commonly used classification techniques require at least one parameter setting, indicating that selecting an optimal parameter setting for defect prediction models is an important experimental design choice.*

## 3 Related Work & Research Questions

Recent research has raised concerns about parameter settings of classification techniques when applied to defect prediction models. For example, Koru *et al.* [42] and Mende *et al.* [93, 94] point out that selecting different parameter settings can impact the performance of defect models. Jiang *et al.* [63] and Tosun *et al.* [146] also point out that the default parameter settings of research toolkits (e.g., R [113], Weka [44], Scikit-learn [106], MATLAB [89]) are suboptimal.

Although prior work suggests that defect prediction models may underperform if they are trained using suboptimal parameter settings, parameters are often left at their default values. Fu *et al.* [35] point out that 80% of the 50 highly-cited defect prediction studies rely on a default setting that is provided by the employed research toolkits (e.g., R or Weka). For example, Mende *et al.* [92] use the default number of decision trees to train a random forest classifier (provided by an R package). Weyuker *et al.* [148] also train defect models using the default setting of C4.5 that is provided by Weka. Jiang *et al.* [62] and Bibi *et al.* [13] use the default value of $k$ for the $k$-nearest neighbours classification technique ($k = 1$). In our prior work (e.g., [39, 138]), we ourselves have also used default classification settings.

In addition, the implementations of classification techniques that are provided by different research toolkits often use different default settings. For example, for the number of decision trees for a random forest classifier, the default varies among settings of 10 for the `bigrf` R package [83], 50 for MATLAB [89], 100 for Weka [44], and 500 for the `randomForest` R package [82]. Moreover, for the number of hidden layers of the neural networks techniques, the default varies among settings of 1 for the `neuralnet` R package [33], 2 for Weka [44] and the `nnet` R package [118], and 10 for MATLAB [89]. Such variations among default settings of different research toolkits may influence conclusions of defect prediction studies [135].

There are many empirical studies in the area of Search-Based Software Engineering (SBSE) [10, 46, 47, 73] that aim to optimize software engineering tasks (e.g., software testing [60]). However, little is known about the impact of automated parameter optimization on the performance of defect prediction models. Thus, we formulate the following research question:

*(RQ1) What is the impact of automated parameter optimization on the performance of defect prediction models?*

Like any form of classifier optimization, automated parameter optimization may increase the risk of *overfitting*, i.e., producing classifiers that are too specialized for the data from which they were trained to apply to other datasets. Such overfitting classifiers may produce unstable performance estimates.

Indeed, recent research voices concerns about the stability of performance estimates that are obtained from classification techniques when applied to defect prediction models. For example, Menzies *et al.* [97] and Mittas *et al.* [98] argue that unstable classification techniques can make replication of defect prediction studies more difficult. Shepperd *et al.* [123], and Jorgensen *et al.* [67] point out that the unstable performance estimates that are produced by classification techniques may introduce bias, which can mislead different research groups to draw erroneous conclusions. Mair *et al.* [88] and Myrtveit *et al.* [101] show that high variance in performance estimates from classification techniques is a critical problem in comparative studies of prediction models. Song *et al.* [128] also show that applying different settings to unstable classification techniques will provide different results.

To investigate whether parameter optimization impacts the performance stability of defect prediction models, we formulate the following research question:

*(RQ2) Does automated parameter optimization increase the performance instability of defect prediction models?*

In addition to being used for prediction, defect models are also used to understand the characteristics of defect-prone modules. For example, Bettenburg *et al.* [12] study the relationship between social interactions and software quality. Shihab *et al.* [126] study the characteristics of high impact and surprising defects. Recent works [74, 90, 91, 99, 144] explore the relationship between modern code review practices and software quality. Such an understanding of defect-proneness is essential to design effective quality improvement plans.

Recent research draws into question the accuracy of insights that are derived from defect prediction models. For example, our recent work shows that issue report mislabelling has a large impact on the interpretation of defect prediction models [138]. To investigate whether default parameter settings impact the interpretation of defect prediction models, we formulate the following research question:

*(RQ3) How much does the interpretation of defect prediction models change when automated parameter optimization is applied?*

Recent research applies the parameter settings of classification techniques that perform well on one dataset to another. For example, Jiang *et al.* [61] also experiment with various settings on one dataset, since the total execution time of a classification technique (i.e., MARS) can take several hours. However, others [39, 71, 129] show that the performance of defect prediction models often depends on the characteristics of the defect datasets that are used to train them. Gao *et al.* [38] also argue that classification techniques are dataset-specific, i.e., classification techniques that work well on one dataset but may not work well on another dataset. Tan *et al.* [133] explore different settings to identify the optimal setting on one dataset and assume that the optimal setting will apply to other datasets. Yet, little is known about whether the optimal parameter settings that are obtained for one dataset are transferable to another. Knowing the transferability of the optimal parameter settings of the most sensitive classification techniques will shed light onto whether automated parameter optimization can be safely omitted (by opting to use these transferable setting instead) without incurring a significant drop in performance. To investigate how well the optimal parameter settings transfer from one dataset to another, we formulate the following research question:

*(RQ4) How well do optimal parameter settings transfer from one dataset to another?*

In addition to the benefits of applying automated parameter optimization, one must also consider the additional computational cost to provide a balanced perspective on the practicality of parameter optimization. Menzies *et al.* [95] show that some classification techniques can be very slow and exploring various settings would incur a large computational cost. Ma *et al.* [86] point out that the computational cost depends on to the size of the training dataset. Hence, we formulate the following research question:

*(RQ5) What is the computational cost of applying automated parameter optimization?*

## 4 Studied Datasets

In this section, we discuss our selection criteria and the studied datasets that satisfy these criteria. In selecting the studied datasets, we identified three important criteria that needed to be satisfied:

TABLE 1: Overview of the examined parameters of the studied classification techniques.

| Family | Family Description | Parameter Name | Parameter Description | Studied classification techniques with their default (in bold) and candidate parameter values. |
|---|---|---|---|---|
| Naive Bayes | Naive Bayes is a probabilistic model that assumes that predictors are independent of each other [75].<br>**Techniques:** Naive Bayes (NB). | Laplace Correction | [N] Laplace correction (0 indicates no correction). | NB={**0**} |
| | | Distribution Type | [L] TRUE indicates a kernel density estimation, while FALSE indicates a normal density estimation. | NB={TRUE, **FALSE**} |
| Nearest Neighbour | Nearest neighbour is an algorithm that stores all available observations and classifies each new observation based on its similarity to prior observations [75].<br>**Techniques:** $k$-Nearest Neighbour (KNN). | #Clusters | [N] The numbers of non-overlapping clusters to produce. | KNN={**1**, 5, 9, 13, 17} |
| Regression | Logistic regression is a technique for explaining binary dependent variables. MARS is a non-linear regression modelling technique [32].<br>**Techniques:** GLM and MARS. | Degree Interaction | [N] The maximum degree of interaction (Friedman's mi). The default is 1, meaning build an additive model (i.e., no interaction terms). | MARS={**1**} |
| Partial Least Squares | Partial Least Squares regression generalizes and combines features from principal component analysis and multiple regression [142].<br>**Techniques:** Generalized Partial Least Squares (GPLS). | #Components | [N] The number of PLS components. | GPLS={**1**, 2, 3, 4, 5} |
| Neural Network | Neural network techniques are used to estimate or approximate functions that can depend on a large number of inputs and are generally unknown [127].<br>**Techniques:** Standard (NNet), Model Averaged (AVNNet), Feature Extraction (PCANNet), Radial Basis Functions (RBF), Multi-layer Perceptron (MLP), Voted-MLP (MLPWeightDecay), and Penalized Multinomial Regression (PMR). | Bagging | [L] Should each repetition apply bagging? | AVNNet={TRUE, **FALSE**} |
| | | Weight Decay | [N] A penalty factor to be applied to the errors function. | MLPWeightDecay, PMR, AVNNet, NNet, PCANNet={**0**, 0.0001, 0.001, 0.01, 0.1}, SVMLinear={**1**} |
| | | #Hidden Units | [N] Numbers of neurons in the hidden layers of the network that are used to produce the prediction. | MLP, MLPWeightDecay, AVNNet, NNet, PCANNet={**1**, 3, 5, 7, 9},<br>RBF={**11**, 13, 15, 17, 19} |
| Discrimination Analysis | Discriminant analysis applies different kernel functions (e.g., linear) to classify a set of observations into predefined classes based on a set of predictors [31].<br>**Techniques:** Linear Discriminant Analysis (LDA), Penalized Discriminant Analysis (PDA), and Flexible Discriminant Analysis (FDA). | Product Degree | [N] The number of degrees of freedom that are available for each term. | FDA={**1**} |
| | | Shrinkage Penalty Coefficient | [N] A shrinkage parameter that is applied to each tree in the expansion (a.k.a., learning rate or step-size reduction). | PDA={**1**, 2, 3, 4, 5} |
| | | #Terms | [N] The number of terms in the model. | FDA={**10**, 20, 30, 40, 50} |
| Rule-based | Rule-based techniques transcribe decision trees using a set of rules for classification [75].<br>**Techniques:** Rule-based classifier (Rule), and Ripper classifier (Ripper). | #Optimizations | [N] The number of optimization iterations. | Ripper={1, **2**, 3, 4, 5} |
| Decision Trees-Based | Decision trees use feature values to classify instances [75].<br>**Techniques:** C4.5-like trees (J48), Logistic Model Trees (LMT), and Classification And Regression Trees (CART). | Complexity | [N] A penalty factor that is applied to the error rate of the terminal nodes of the tree. | CART={0.0001, 0.001, **0.01**, 0.1, 0.5} |
| | | Confidence | [N] The confidence factor that are used for pruning (smaller values incur more pruning). | J48={**0.25**} |
| | | #Iterations | [N] The number of iterations. | LMT={**1**, 21, 41, 61, 81} |
| SVM | Support Vector Machines (SVMs) use a hyperplane to separate two classes (i.e., defective or not) [75].<br>**Techniques:** SVM with Linear kernel (SVMLinear), and SVM with Radial basis function kernel (SVMRadial). | Sigma | [N] The width of the Gaussian kernels. | SVMRadial={0.1, 0.3, **0.5**, 0.7, 0.9} |
| | | Cost | [N] A penalty factor to be applied to the number of errors. | SVMRadial={0.25, 0.5, **1**, 2, 4}, SVMLinear={**1**} |
| Bagging | Bagging methods combine different base classifiers together to classify instances [154].<br>**Technique:** Random Forest (RF), Bagged CART (BaggedCART) | #Trees | [N] The number of classification trees. | RF={**10**, 20, 30, 40, 50} |
| Boosting | Boosting performs multiple iterations, each with different example weights, and makes predictions using voting of classifiers [18].<br>**Techniques:** Gradient Boosting Machine (GBM), Adaptive Boosting (AdaBoost), Generalized linear and Additive Models Boosting (GAMBoost), Logistic Regression Boosting (LogitBoost), eXtreme Gradient Boosting Tree (xGBTree), and C5.0. | #Boosting Iterations | [N] The numbers of iterations that are used to construct models. | C5.0={**1**, 10, 20, 30, 40}, GAMBoost={**50**, 100, 150, 200, 250}, LogitBoost={**11**, 21, 31, 41, 51}, GBM,xGBTree={50, **100**, 150, 200, 250} |
| | | #Trees | [N] The numbers of classification trees. | AdaBoost={**50**, 100, 150, 200, 250} |
| | | Shrinkage | [N] A shrinkage factor that is applied to each tree in the expansion (a.k.a., learning rate or step-size reduction). | GBM={**0.1**}, xGBTree={**0.3**} |
| | | Max Tree Depth | [N] The maximum depth per tree. | AdaBoost, GBM, xGBTree={**1**, 2, 3, 4, 5} |
| | | Min. Terminal Node Size | [N] The minimum terminal nodes in the trees. | GBM={**10**} |
| | | Winnow | [L] Should predictor winnowing (i.e feature selection) be applied? | C5.0={**FALSE**, TRUE} |
| | | AIC Prune? | [L] Should pruning using stepwise feature selection be applied? | GAMBoost={**FALSE**, TRUE} |
| | | Model Type | [F] Either use trees for the predicted class or rules for model confidence values. | C5.0={**rules**, tree} |

[N] denotes a numeric value; [L] denotes a logical value; [F] denotes a factor value.

The default values are shown in bold typeface and correspond to the default values of the Caret R package.

### Criterion 1 — Publicly-available defect datasets from different corpora

Our recent work [140] highlights that the tendency of researchers to reuse experimental components (e.g., datasets, metrics, and classifiers) can introduce a bias in the reported results. Song *et al.* [129] and Ghotra *et al.* [39, 40] also show that the performance of defect prediction models can be impacted by the dataset from which they are trained. To combat potential bias in our conclusions and to foster replication of our experiments, we choose to train our defect prediction models using datasets from different corpora and domains that are hosted in publicly-available data repositories. To satisfy criterion 1, we began our study using 101 publicly-available defect datasets: 76 datasets from the Tera-PROMISE Repository,[3] 12 clean NASA datasets as provided by Shepperd *et al.* [124], 5 datasets as provided by Kim *et al.* [72, 149], 5 datasets as provided by D'Ambros *et al.* [23, 24], and 3 datasets as provided by Zimmermann *et al.* [155].

### Criterion 2 — Dataset that has a low risk of overfitting

Mende *et al.* [93] show that models that are trained using small datasets may produce unstable performance estimates. An influential characteristic in the performance of a classification technique is the number of *Events Per Variable* (EPV) [107, 141], i.e., the ratio of the number of occurrences of the least frequently occurring class of the dependent variable (i.e., the number of defective modules) to the number of independent variables that are used to train the model (i.e., the numbers of variables). For example, given a defect dataset of 1,000 modules with 10 software metrics and a defective ratio of 10%, the dataset has an EPV value of 10 (i.e., = 100 defective modules / 10 variables). Larger EPV values indicate the lower risk of producing unstable results due to overfitting. Our recent work shows that defect prediction models that are trained using datasets with low EPV values are especially susceptible to unstable results [141]. To mitigate this risk, we choose to study datasets that have an EPV above 10, as suggested by Peduzzi *et al.* [107]. To satisfy criterion 2, we exclude the 78 datasets that we found to have an EPV value that is below 10.

### Criterion 3 — Dataset that produced non-overly optimistic model performance

Classification techniques that are trained on imbalanced data often favour the majority class. When defective modules are the majority class, defect prediction models are likely to produce overly optimistic performance values. Moreover, for a defect dataset with a high defective ratio, a ZeroR classifier may suffice. ZeroR is a simple classification technique, which predicts that all instances in the testing dataset are of the majority class from the training dataset [55]. To satisfy criterion 3, we exclude an additional 5 datasets because they have a defective rate above 50%.

### Datasets that satisfy our three criteria

Table 2 provides an overview of the 18 datasets that satisfy our analysis criteria. To strengthen the generalizability of our results, the studied datasets include proprietary and open source systems of varying size and domain.

3. http://openscience.us/repo/

TABLE 2: An overview of the studied datasets.

| Domain | Dataset | Defective Rate | #Modules | #Metrics | EPV | Granularity |
|---|---|---|---|---|---|---|
| NASA | JM1[1] | 21% | 7,782 | 21 | 80 | Method |
| | PC5[1] | 28% | 1,711 | 38 | 12 | Method |
| Proprietary | Prop-1[2] | 15% | 18,471 | 20 | 137 | Class |
| | Prop-2[2] | 11% | 23,014 | 20 | 122 | Class |
| | Prop-3[2] | 11% | 10,274 | 20 | 59 | Class |
| | Prop-4[2] | 10% | 8,718 | 20 | 42 | Class |
| | Prop-5[2] | 15% | 8,516 | 20 | 65 | Class |
| Apache | Camel 1.2[2] | 36% | 608 | 20 | 11 | Class |
| | Xalan 2.5[2] | 48% | 803 | 20 | 19 | Class |
| | Xalan 2.6[2] | 46% | 885 | 20 | 21 | Class |
| Eclipse | Platform 2.0[3] | 14% | 6,729 | 32 | 30 | Files |
| | Platform 2.1[3] | 11% | 7,888 | 32 | 27 | File |
| | Platform 3.0[3] | 15% | 10,593 | 32 | 49 | File |
| | Debug 3.4[4] | 25% | 1,065 | 17 | 15 | Method |
| | SWT 3.4[4] | 44% | 1,485 | 17 | 38 | Method |
| | JDT[5] | 21% | 997 | 15 | 14 | Class |
| | Mylyn[5] | 13% | 1,862 | 15 | 16 | Class |
| | PDE[5] | 14% | 1,497 | 15 | 14 | Class |

[1]Provided by Shepperd *et al.* [124].
[2]Provided by Jureczko *et al.* [68].
[3]Provided by Zimmermann *et al.* [155].
[4]Provided by Kim *et al.* [72, 149].
[5]Provided by D'Ambros *et al.* [23].

## 5 Case Study Approach

In this section, we describe the design of the case study that we perform in order to address our five research questions. Figure 1 provides an overview of the approach that we apply to each studied dataset. We describe each step in our approach below.

### 5.1 Generate Bootstrap Sample

In order to ensure that our conclusions are robust, we use the out-of-sample bootstrap validation technique [28, 141], which leverages aspects of statistical inference [29]. The out-of-sample bootstrap is made up of two steps:

**(Step 1)** A bootstrap sample of size $N$ is randomly drawn with replacement from an original dataset, which is also of size $N$.

**(Step 2)** A model is trained using the bootstrap sample and tested using the rows that do not appear in the bootstrap sample. On average, 36.8% of the rows will not appear in the bootstrap sample, since the sample is drawn with replacement [28].

The out-of-sample bootstrap process is repeated 100 times, and the average out-of-sample performance is reported as the performance estimate.

Unlike the ordinary bootstrap, the out-of-sample bootstrap technique fits models using the bootstrap samples, but rather than testing the model on the original sample, the model is instead tested using the rows that do not appear in the bootstrap sample [141]. Thus, the training and testing corpora do not share overlapping observations.

Unlike $k$-fold cross-validation, the out-of-sample bootstrap technique fits models using a dataset that is of equal length to the original dataset. Cross-validation splits the data into $k$ equal parts, using $k$ - 1 parts for fitting the model, setting aside 1 fold for testing. The process is repeated $k$ times, using a different part for testing each time. However, Mende *et al.* [93] point out that the scarcity of defective modules in the small testing corpora of 10-fold cross validation may produce biased and unstable results.

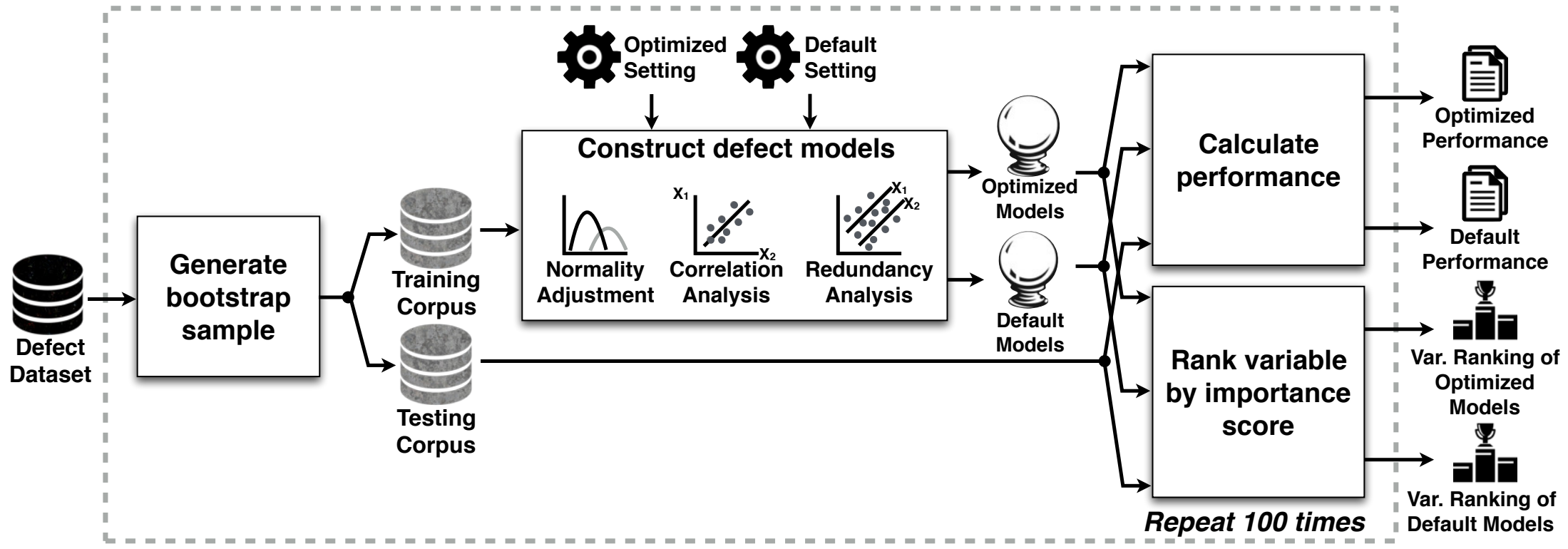

Fig. 1: An overview of our case study approach.

Prior studies have also shown that 10-fold cross validation can produce unstable results for small samples [17]. On the other hand, our recent research demonstrates that the out-of-sample bootstrap tends to produce the least biased and most stable performance estimates [141]. Moreover, the out-of-sample bootstrap is recommended for highly-skewed datasets [49], as is the case in our defect prediction datasets.

### 5.2 Identify the Optimal Settings

To address the first six research questions, we use the grid-search optimized parameter settings as suggested by the `train` function of the `caret` R package [78]. Caret suggests candidate settings for each of the studied classification techniques, which can be checked using the `getModelInfo` function of the `caret` R package [78]. A key benefit of applying a grid-search optimization technique is that the experimental design is highly controlled—i.e., the candidate parameter settings of the grid search technique do not vary for each dataset and bootstrap sample, while the candidate parameter settings of other advanced parameter optimization techniques often vary, leading to substantially more complex experimental settings. Figure 2 provides an overview of the grid-search parameter optimization process. The optimization process is made up of three steps.

**(Step 1)** **Generate candidate parameter settings:** The `train` function will generate candidate parameter settings based on a given budget threshold (i.e., the `tuneLength`) for evaluation. The budget threshold indicates the number of different values to be evaluated for each parameter. As suggested by Kuhn [76], we use a budget threshold of 5. For example, the number of boosting iterations of the C5.0 classification technique is initialized to 1 and is increased by 10 until the number of candidate settings reaches the budget threshold (e.g., 1, 11, 21, 31, 41). Table 1 shows the candidate parameter settings for each of the examined parameters. The default settings are shown in bold typeface.

**(Step 2)** **Evaluate candidate parameter settings:** Caret evaluates all of the potential combinations of the candidate parameter settings. For example, if a classification technique accepts 2 parameters with 5 candidate parameter settings for each, Caret will explore all 25 potential combinations of parameter settings. We use 100 repetitions of the out-of-sample bootstrap to estimate the performance of classifiers that are trained using each of the candidate parameter settings. For each candidate parameter setting, a classifier is fit to a sample of the training corpus and we estimate the performance of a model using those rows in the training corpus that do not appear in the sample that was used to trained the classifier.

**(Step 3)** **Identify the optimal setting:** Finally, the performance estimates are used to identify the most optimal parameter settings.

### 5.3 Construct Defect Prediction Models

To measure the impact of automated parameter optimization on defect prediction models, we train our models using the optimized and default settings. To ensure that the training and testing corpora have similar characteristics, we do not re-balance nor do we re-sample the training data.

**Normality Adjustment**. Analysis of the distributions of our independent variables reveals that they are right-skewed. As suggested by previous research [4, 11, 15], we mitigate this skew by log-transforming each independent variable using $ln(x + 1)$ prior to training our models.

**Correlation Analysis**. Highly correlated independent variables may interfere with each other when a model is being interpreted. Indeed, our recent work [65, 136, 140] has demonstrated that collinearity and multicollinearity issues can artificially inflate (or deflate) the impact of software metrics when interpreting defect prediction models. Jiarpakdee *et al.* [66] also point out that 10%-67% of metrics of publicly-available defect datasets are redundant. Thus, we perform correlation and redundancy analyses prior to training our defect prediction models. We measure the correlation between explanatory variables using Spearman rank correlation tests ($\rho$), as it relaxes the non-normality assumption of the data. We then use a variable clustering analysis [120] to construct a hierarchical overview of the correlation and remove explanatory variables with a high correlation. We select $|\rho| = 0.7$ as a threshold for removing highly correlated variables [70]. We perform this analysis iteratively until all clusters of surviving variables have $|\rho| < 0.7$.

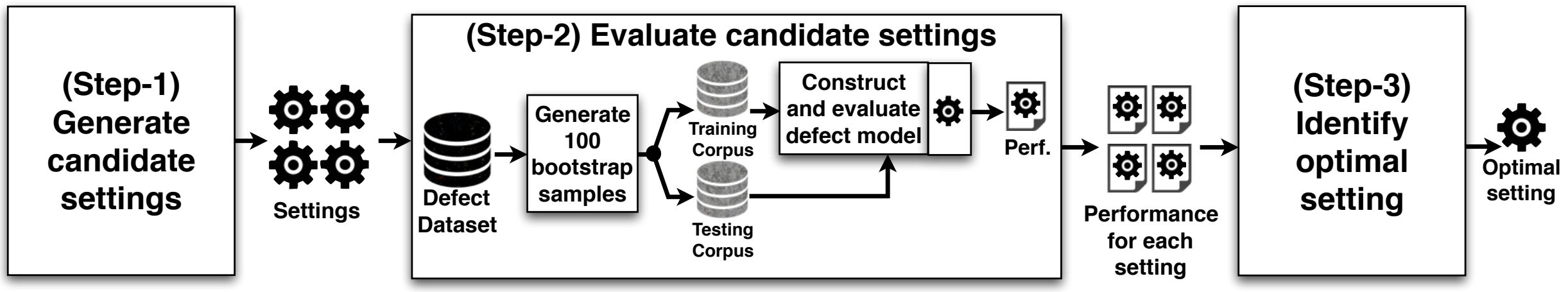

Fig. 2: An overview of grid-search parameter optimization.

**Redundancy Analysis**. While correlation analysis reduces collinearity among our variables, it does not detect all of the *redundant variables*, i.e., variables that do not have a unique signal with respect to the other variables. Redundant variables will interfere with each other, distorting the modelled relationship between the explanatory variables and the outcome. Therefore, we remove redundant variables prior to constructing our defect prediction models. In order to detect redundant variables, we fit preliminary models that explain each variable using the other explanatory variables. We use the $R^2$ value of the preliminary models to measure how well each variable is explained by the other variables.

We use the implementation of this technique as provided by the `redun` function of the `rms` R package [50]. The variable that is most well-explained by the other variables is iteratively dropped until either: (1) no preliminary model achieves an $R^2$ above a cutoff threshold (for this paper, we use the default threshold of 0.9), or (2) removing a variable would make a previously dropped variable no longer explainable, i.e., its preliminary model will no longer achieve an $R^2$ exceeding the threshold.

### 5.4 Measuring the performance of the models

We apply the defect prediction models that we train using the training corpus to the testing corpus in order to measure their performance. Prior work focused on a limited number of performance measures. For example, our prior work [139] study only one threshold-independent measures (i.e., AUC). Fu *et al.* [35] study three threshold-dependent measures (i.e., Precision, Recall, and F-measure). To combat any potential bias and extend the generalization of our findings, we apply both threshold-dependent and threshold-independent performance measures to quantify the performance of our models. First, we include the three traditional performance measures for defect prediction studies, i.e., Precision (i.e,. Positive Predicted Values), Recall (i.e., Probability of Detection, True Positive rate, Sensitivity, PD), and F-measure, as previously used by Fu *et al.* [35]. Since Zhang *et al.* [151, 152] argue that precision and recall can be unstable when datasets contain a low percentage of defective ratios, we include Gmean, Gmeasure, and True Negative rate (i.e,. Specificity). As suggested by Menzies *et al.* [95], we include False Positive rate (i.e, Probability of False Alarm, PF). As suggested by Zhang *et al.* [153] and Tosun *et al.* [145], we include Balance measure. As suggested by Shepperd *et al.* [123], we include Matthews Correlation Coefficient.

Prior studies argued that threshold-dependent performance metrics (e.g., precision and recall) are problematic because they: (1) depend on an arbitrarily-selected threshold [1, 5, 81, 114, 125, 137] and (2) are sensitive to imbalanced data [25, 51, 85, 93, 133, 137]. Thus, we also include three threshold-independent performance measures to quantify the performance of our defect prediction models (i.e., AUC, Brier, and LogLoss). Arisholm *et al.* [5] argue that traditional performance measures do not consider the required inspection cost. Therefore, recent studies [69, 114–116] recommend the use of the area under the cost effectiveness curve (AUCEC). However, the AUCEC measure is not currently compatible with the Caret implementation. Caret measures the model performance based on a `summaryFunction` function that only takes the observed and predicted values. Hence, we are unable to compute the AUCEC measure. Nonetheless, we plan to explore the impact of automated parameter optimization on the AUCEC measure in the future.

Table 3 provide the definitions and descriptions of our 9 threshold-dependent and 3 threshold-independent performance measures. In total, we study 12 performance measures that are commonly-used in defect prediction studies. We describe and discuss each performance measure below.

#### *5.4.1 Threshold-Dependent Performance Measures*

In order to calculate the threshold-dependent performance measures, these probabilities are transformed into a binary classification (defective or clean) using the default threshold value of 0.5, i.e., if a module has a predicted probability above 0.5, it is considered defective; otherwise, the module is considered clean. Using the default threshold of 0.5, we compute nine threshold-dependent performance measures.

#### *5.4.2 Threshold-Independent Performance Measures*

First, we use the *Area Under the receiver operator characteristic Curve (AUC)* to measure the discriminatory power of our models, as suggested by recent research [26, 49, 58, 81, 130, 131]. The AUC is a threshold-independent performance metric that measures a classifier's ability to discriminate between defective and clean modules (i.e., do the defective modules tend to have higher predicted probabilities than clean modules?). AUC is computed by measuring the area under the curve that plots the true positive rate against the false positive rate, while varying the threshold that is used to determine whether a file is classified as defective or not. Values of AUC range between 0 (worst performance), 0.5 (random guessing performance), and 1 (best performance).

Second, we use the *Brier score* [20, 119] to measure the distance between the predicted probabilities and the

TABLE 3: The definitions and descriptions of our 9 threshold-depending and 3 threshold-independent performance measures. In total, we study 12 performance measures.

| Measure | Definition | Description |
|---|---|---|
| ***Threshold-Dependent Performance Measures*** | | |
| Precision *(Positive Predicted Values)* | $P = \frac{TP}{TP+FP}$ | Proportion of modules that are correctly classified as defective. |
| Recall *(Probability of Detection, True Positive rate, Sensitivity)* | $R = \frac{TP}{TP+FN}$ | Proportion of defective modules that are correctly classified. |
| F-Measure | $F = 2 \times \frac{P \times R}{P+R}$ | Harmonic mean of Precision and Recall. |
| G-mean | $Gmean = \sqrt{TP_{rate} \times TN_{rate}}$ | Geometric mean of True Positive rate and True Negative rate. |
| G-measure | $Gmeasure = \frac{2 \times pd \times (1-fpr)}{pd+(1-fpr)}$ | Harmonic mean of recall and specificity. |
| Balance | $1 - \sqrt{\frac{(0-\text{pf})^2+(1+\text{pd})^2)}{2}}$ | Proportion of clean modules that are misclassified. |
| Matthews Correlation Coefficient | $MCC = \frac{TP \times TN - FP \times FN}{\sqrt{(TP+FP)(TP+FN)(TN_FP)(TN+FN)}}$ | Discrimination power. |
| True Negative rate *(Specificity)* | $TN_{rate} = \frac{TN}{TN+FP}$ | Proportion of clean modules that are correctly classified. |
| False Positive rate *(Probability of False Alarm)* | $FP_{rate} = \frac{FP}{FP+TN}$ | Proportion of clean modules that are misclassified. |
| ***Threshold-Independent Performance Measures*** | | |
| AUC | See Section 5.4.2. | Discriminative power. |
| Brier | $\frac{1}{N}\sum_{i=1}^{N}(p_i - y_i)^2$ | The distance between the predicted probabilities and the outcome. |
| LogLoss | $-\frac{1}{N}\sum_{i=1}^{N}(y_i log(p_i) + (1-y_i)log(1-p_i))$ | Classification loss function. |

outcome, as suggested by Tantithamthavorn *et al.* [141]. The Brier score is calculated as:

$$\text{B} = \frac{1}{N}\sum_{i=1}^{N}(p_i - y_i)^2 \quad (1)$$

where $p_i$ is the predicted probability, $y_i$ is the outcome for module $t$ encoded as 0 if module $t$ is clean and 1 if it is defective, and $N$ is the total number of modules. The Brier score ranges from 0 (best classifier performance) to 1 (worst classifier performance), where a brier score of 0.25 is a random-guessing performance.

Third, we use LogLoss since it (1) penalizes classifiers that are accurate about a false classification [64], (2) is threshold-independent, and (3) is a standard performance measure of the Kaggle competition.[4] LogLoss is computed using the error between predicted probabilities ($p_i$) and outcomes ($y_i$) as follows:

$$logloss = -\frac{1}{N}\sum_{i=1}^{N}(y_i log(p_i) + (1-y_i)log(1-p_i)) \quad (2)$$

A perfect classifier should have a `logLoss` of precisely zero, while less ideal classifiers have progressively larger values of `logLoss`.

## 5.5 Rank Variables by Importance Score

To identify the most important variables, we compute variable importance for each variable in our models. To do so, we introduce a generic variable importance score that can be applied to any classifier, that is derived from Breiman's Variable Importance Score [19]. Figure 3 provides an overview of the calculation of our variable importance measurement to generate the ranks of important variables for each of the optimized-setting and default-setting models.

### 5.5.1 Generic Variable Importance Score

The calculation of our variable importance score consists of 2 steps for each variable.

**(Step 1)** For each testing dataset, we first randomly permute the values of the variable, producing a dataset with that one variable permuted, while all other variables as is.

**(Step 2)** We then compute the difference in the misclassification rates of defect prediction models that are trained using the clean datasets and the datasets with the randomly-permuted variables. The larger the difference, the greater the importance of that particular variable.

We repeat the Steps 1 and 2 for each variable in order to produce a variable important score across all the variables. Since the experiment is repeated 100 times, each variable will have several variable importance scores (i.e., one score for each of the repetitions).

### 5.5.2 Ranking Variables

To study the impact that the studied variables have on our models, we apply the Scott-Knott Effect Size Difference

4. https://www.kaggle.com/wiki/LogLoss

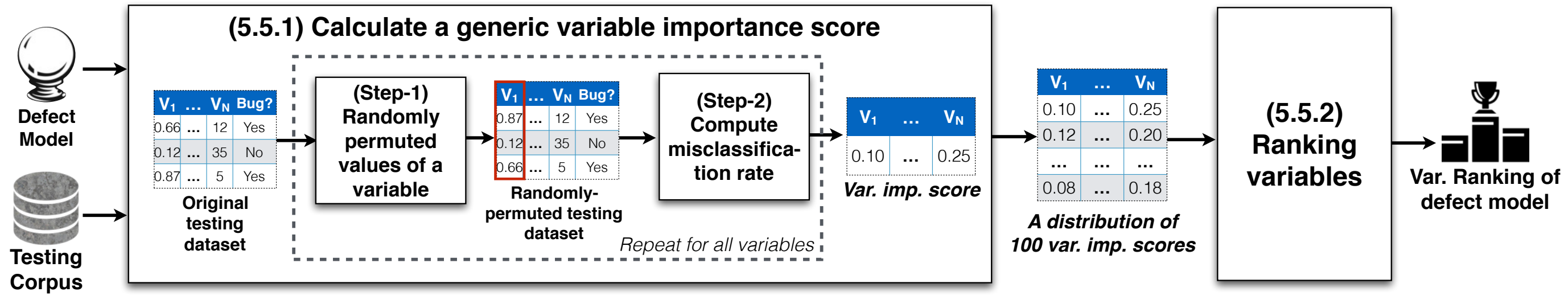

Fig. 3: An overview of our generic variable importance calculation that can be applied to any classification techniques.

(ESD) test (v2.0) [134, 141]. The Scott-Knott ESD test also overcomes the confounding factor of overlapping groups that are produced by other post-hoc tests [39, 98], such as Nemenyi's test [102]. In particular, Nemenyi's test produces overlapping groups of treatments (e.g., variables), implying that there exists no statistically significant difference among the defect prediction models that are trained using many different variables [39]. We use the implementation of the Scott-Knott ESD test of the `sk_esd` function that is provided by the `ScottKnottESD` R package [134]. The Scott-Knott ESD test clusters variables according to statistically significant differences in their mean variable importance scores ($\alpha = 0.05$). The Scott-Knott ESD test ranks each variable exactly once, but several variables may appear within one rank. Finally, we produce rankings of variables in the optimized and default-setting models. Thus, each variable has a rank for each type of model.

We discuss the implementation details of the Scott-Knott ESD test (v2.0) below.

#### 5.5.3 *The Scott-Knott ESD test (v2.0)*

The Scott-Knott ESD test (v2.0) is a mean comparison approach that leverages a hierarchical clustering to partition a set of treatment means (e.g., means of variable importance scores, means of model performance) into statistically distinct groups *with non-negligible difference*.

Unlike the original Scott-Knott test [59], the Scott-Knott ESD test (v2.0) is an alternative approach of the Scott-Knott test [59] that considers the magnitude of the difference (i.e., effect size) of treatment means with-in a group and between groups. Therefore, the Scott-Knott ESD test (v2.0) produces the ranking of treatment means while ensuring that (1) the magnitude of the difference for all of the treatments in each group is negligible; and (2) the magnitude of the difference of treatments between groups is negligible.

Unlike the earlier version of the Scott-Knott ESD test [141] that post-processes the groups that are produced by the Scott-Knott test, the Scott-Knott ESD test (v2.0) check the magnitude of the difference throughout the clustering process by merging pairs of statistically distinct groups that have a negligible difference for all of the treatments of those two groups.

The mechanism of the Scott-Knott ESD test (v2.0) is made up of 2 steps:

**(Step 1)** **Find a partition that maximizes treatment means between groups.** We begin by sorting the treatment means. Then, following the original Scott-Knott test [59], we compute the sum of squares between groups (i.e., a dispersion measure of data points) to identify a partition that maximizes treatment means between groups.

**(Step 2)** **Splitting into two groups or merging into one group.** Instead of using a likelihood ratio test and a $\chi^2$ distribution as a splitting and merging criterion (i.e., a hypothesis testing of the equality of all treatment means) [59], we analyze the magnitude of the difference for each pair for all of the treatment means of the two groups. If there is any one pair of treatment means of two groups are non-negligible, we split into two groups. Otherwise, we merge into one group. We use the Cohen [22] effect size — an effect size estimate based on the difference between the two means divided by the standard deviation of the two treatment means ($d = \frac{\bar{x}_1 - \bar{x}_2}{s.d.}$). The magnitude is assessed using the thresholds provided by Cohen [22] :

$$\text{effect size} = \begin{cases} \textit{negligible} & \text{if Cohen's } d \leq 0.2 \\ \textit{small} & \text{if } 0.2 < \text{Cohen's } d \leq 0.5 \\ \textit{medium} & \text{if } 0.5 < \text{Cohen's } d \leq 0.8 \\ \textit{large} & \text{if } 0.8 < \text{Cohen's } d \end{cases}$$

The implementation of the Scott-Knott ESD test (v2.0) is available at the `sk_esd` function of the `ScottKnottESD` R package [134].

## 6 Case Study Results

In this section, we present the results of our case study with respect to our five research questions.

### (RQ1) What is the impact of automated parameter optimization on the performance of defect prediction models?

**Approach**. To address RQ1, for each performance measure, we start with the performance distribution of the 26 classification techniques that require at least one parameter setting (see Section 2). For each classification technique, we compute the difference in the performance of classifiers that are trained using default settings and optimized parameter settings. We then use boxplots to present the distribution of the performance difference for each of the 18 studied datasets. We apply a Mann-Whitney U test ($\alpha = 0.05$) to assess if the distributions of the performance of classifiers that are trained using default settings and optimized parameter settings are statistically significant. To quantify

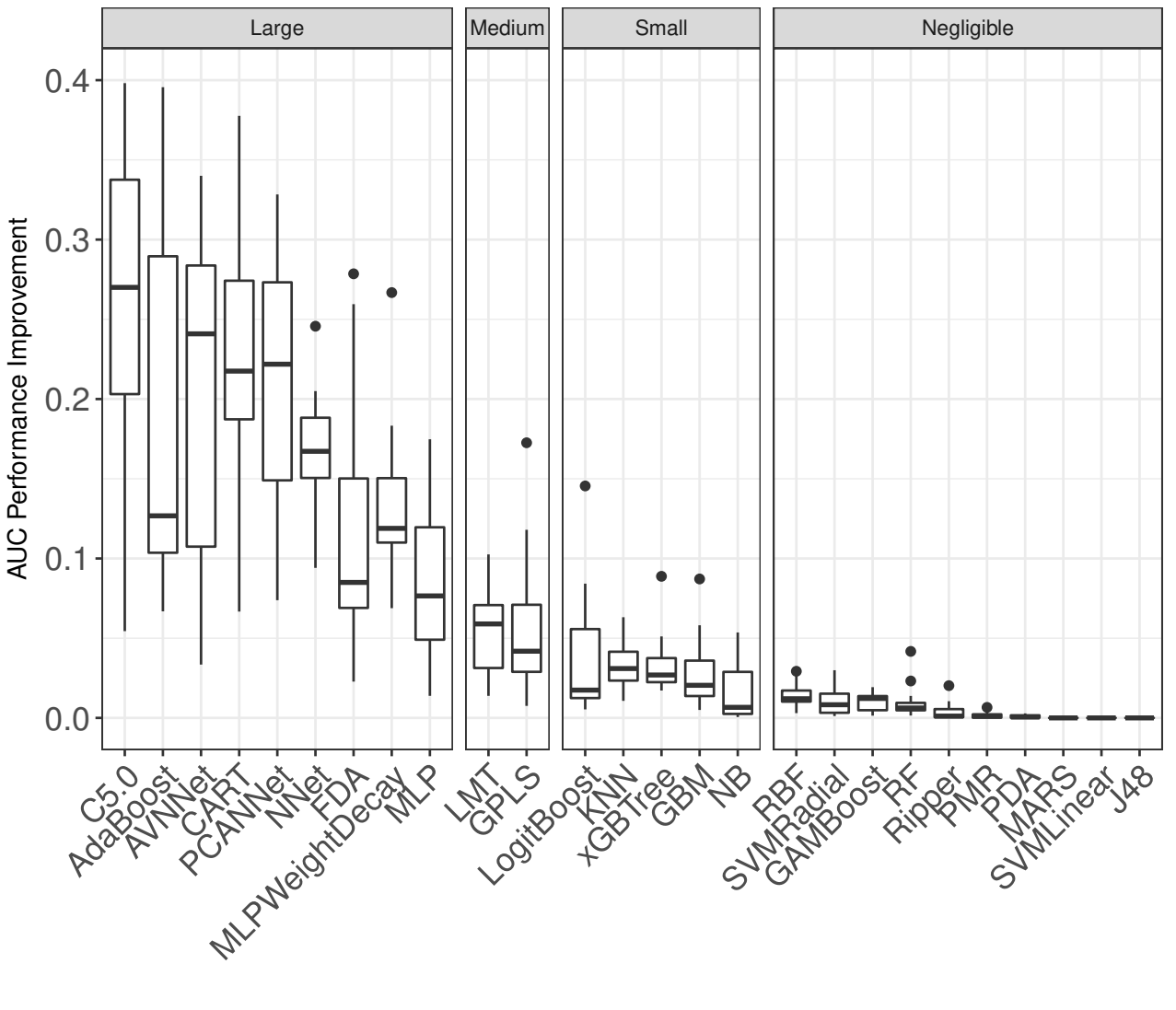

(a) AUC Performance Improvement.

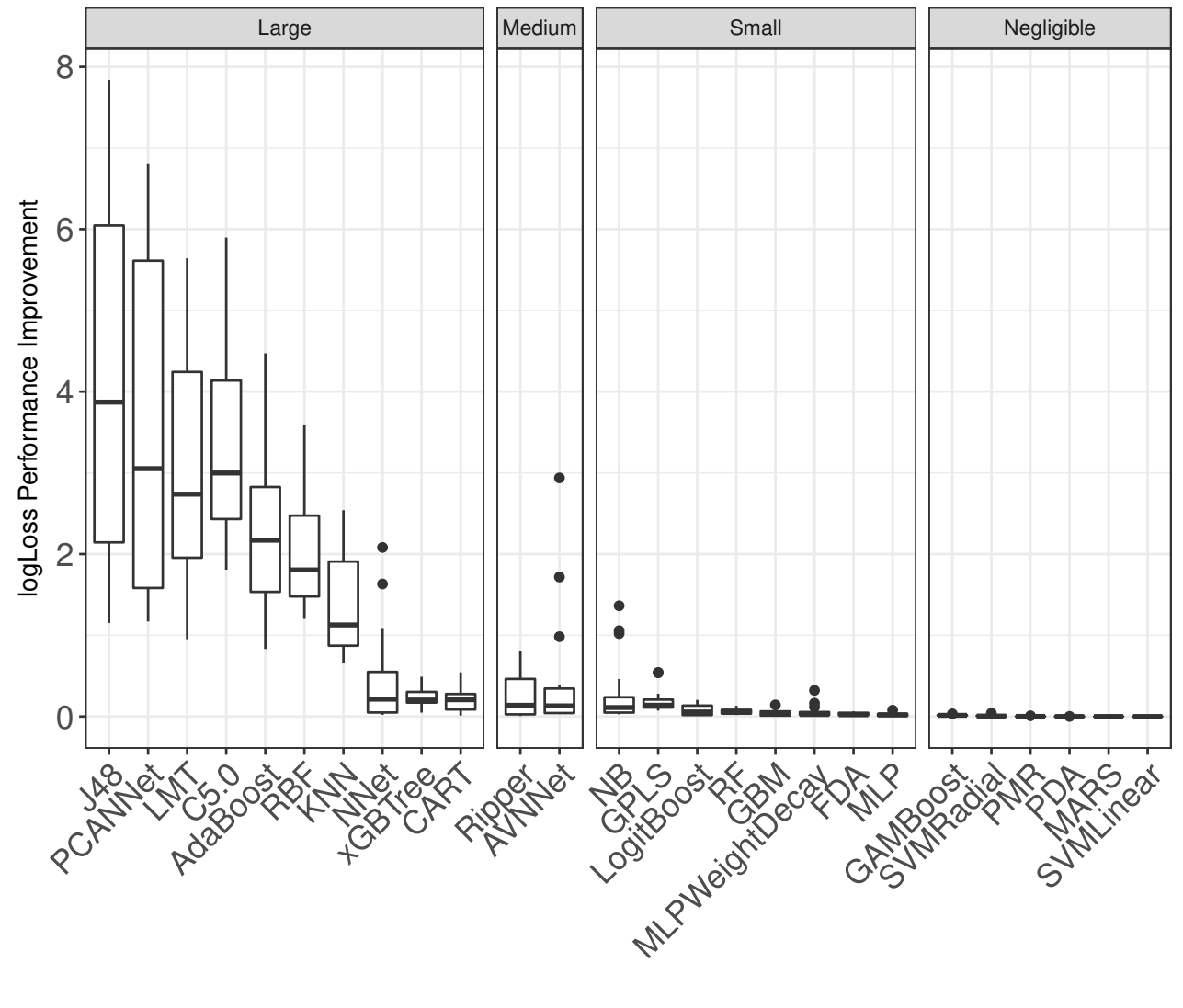

(b) LogLoss Performance Improvement

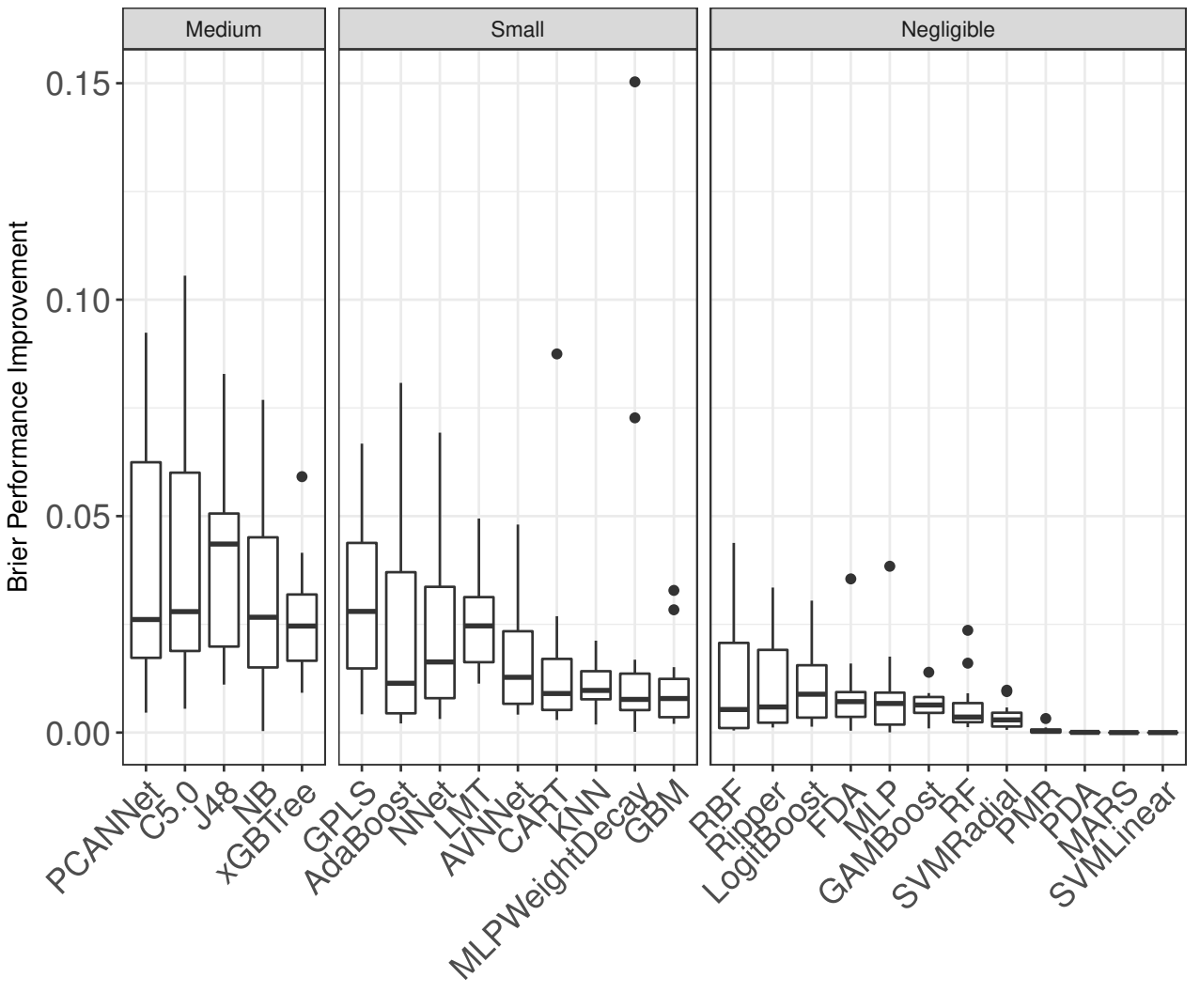

(c) Brier Performance Improvement

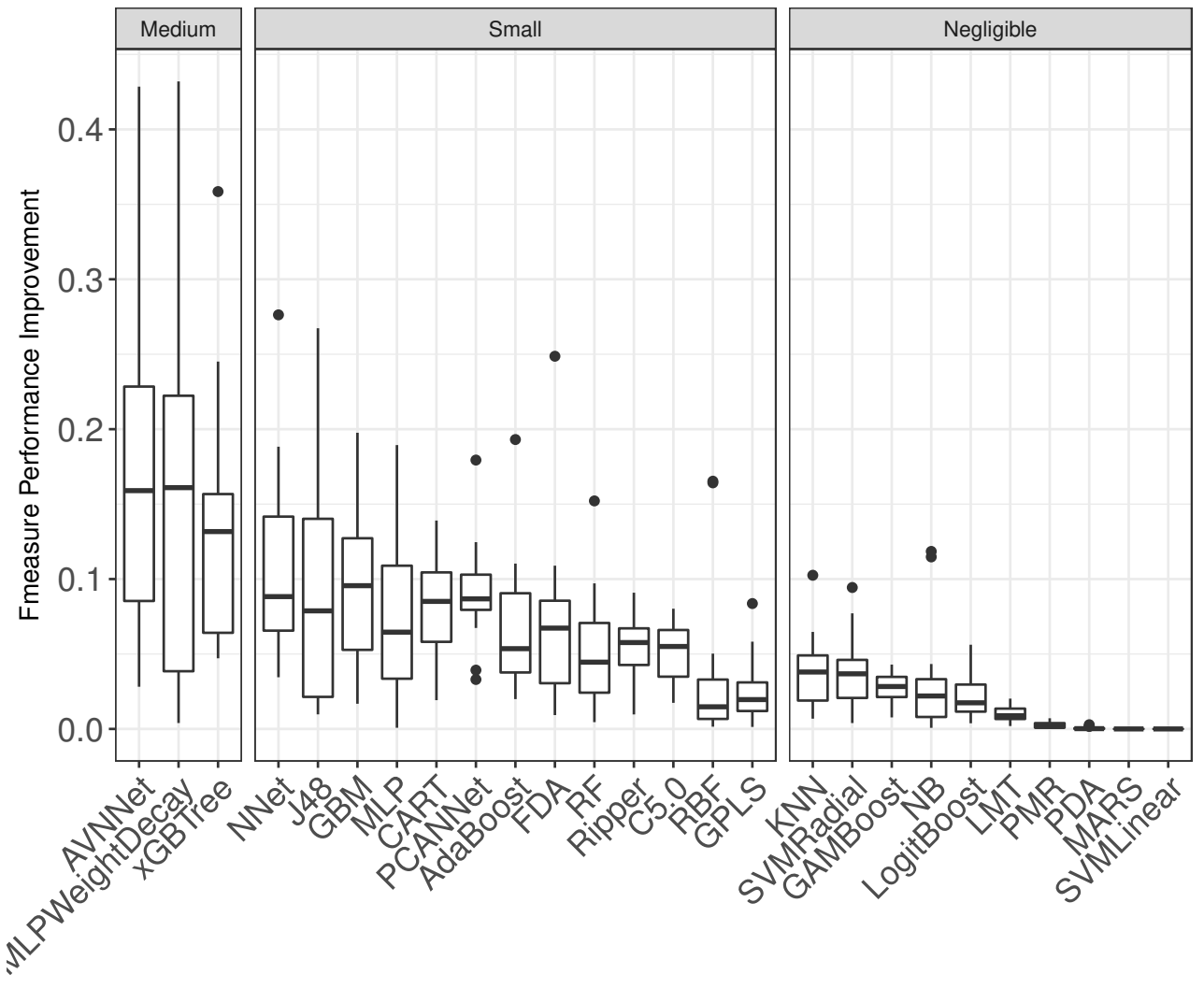

(d) F-Measure Performance Improvement

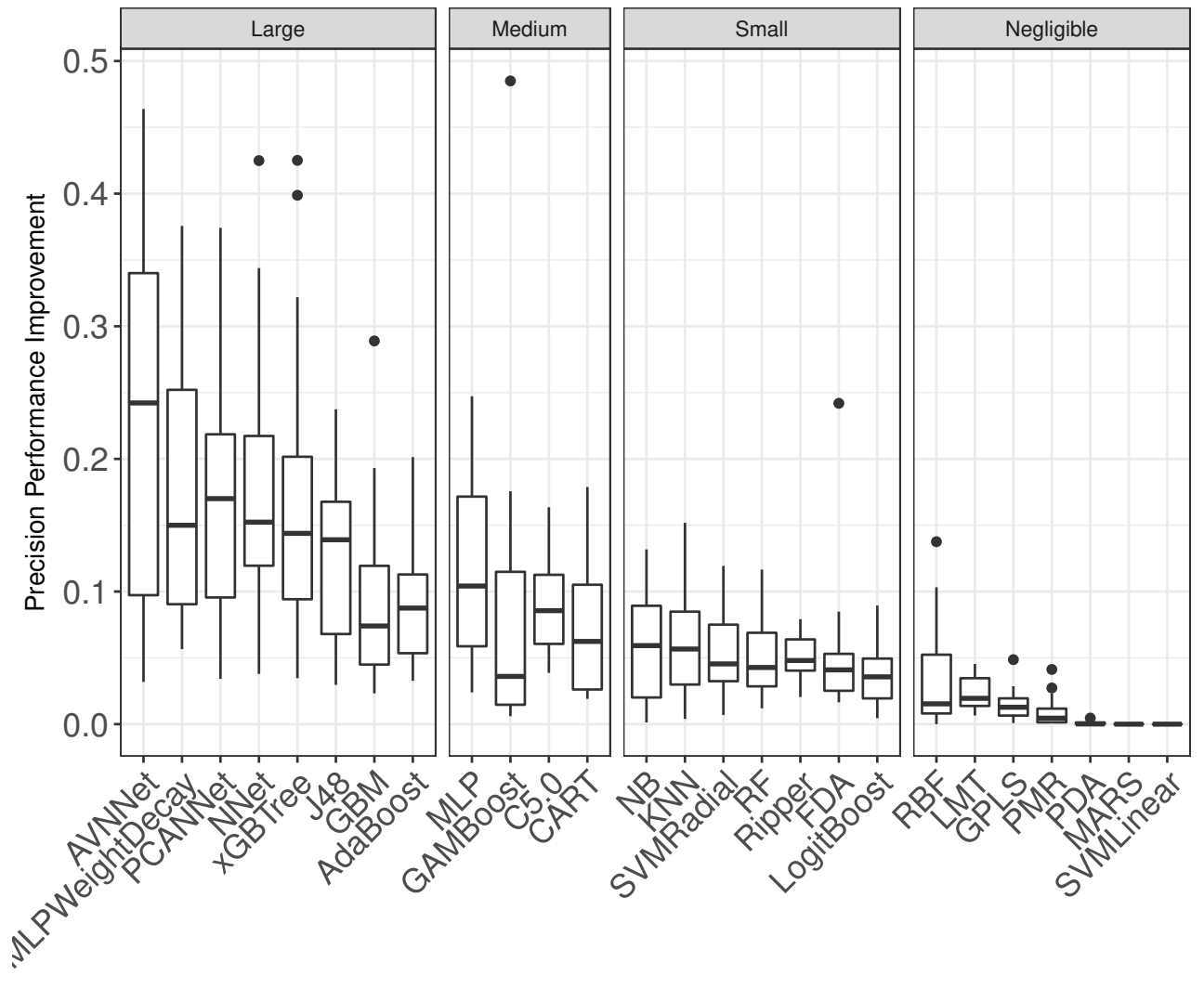

(e) Precision Performance Improvement

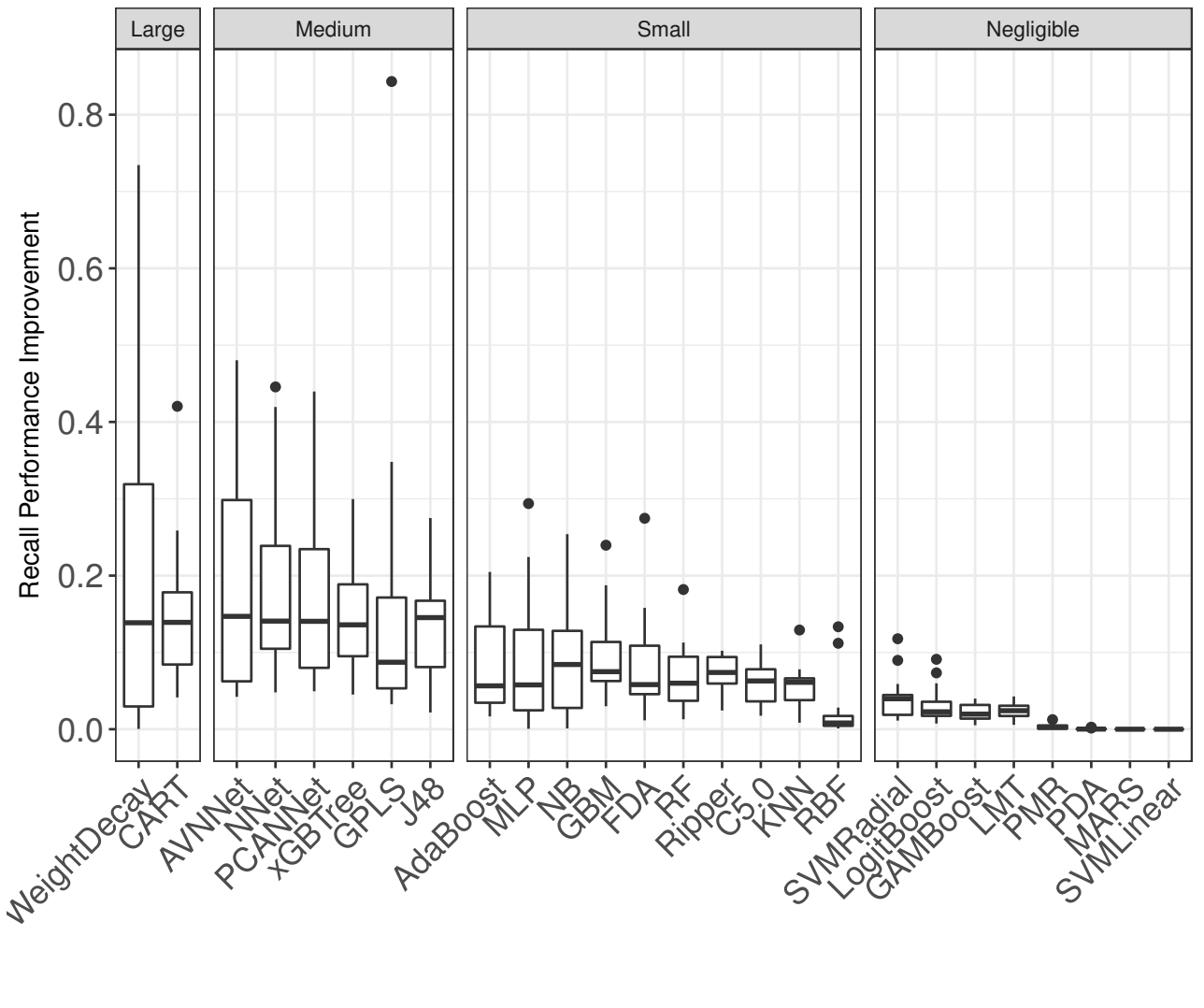

(f) Recall Performance Improvement

Fig. 4: The performance improvement and its Cohen's $d$ effect size for each of the studied classification techniques.

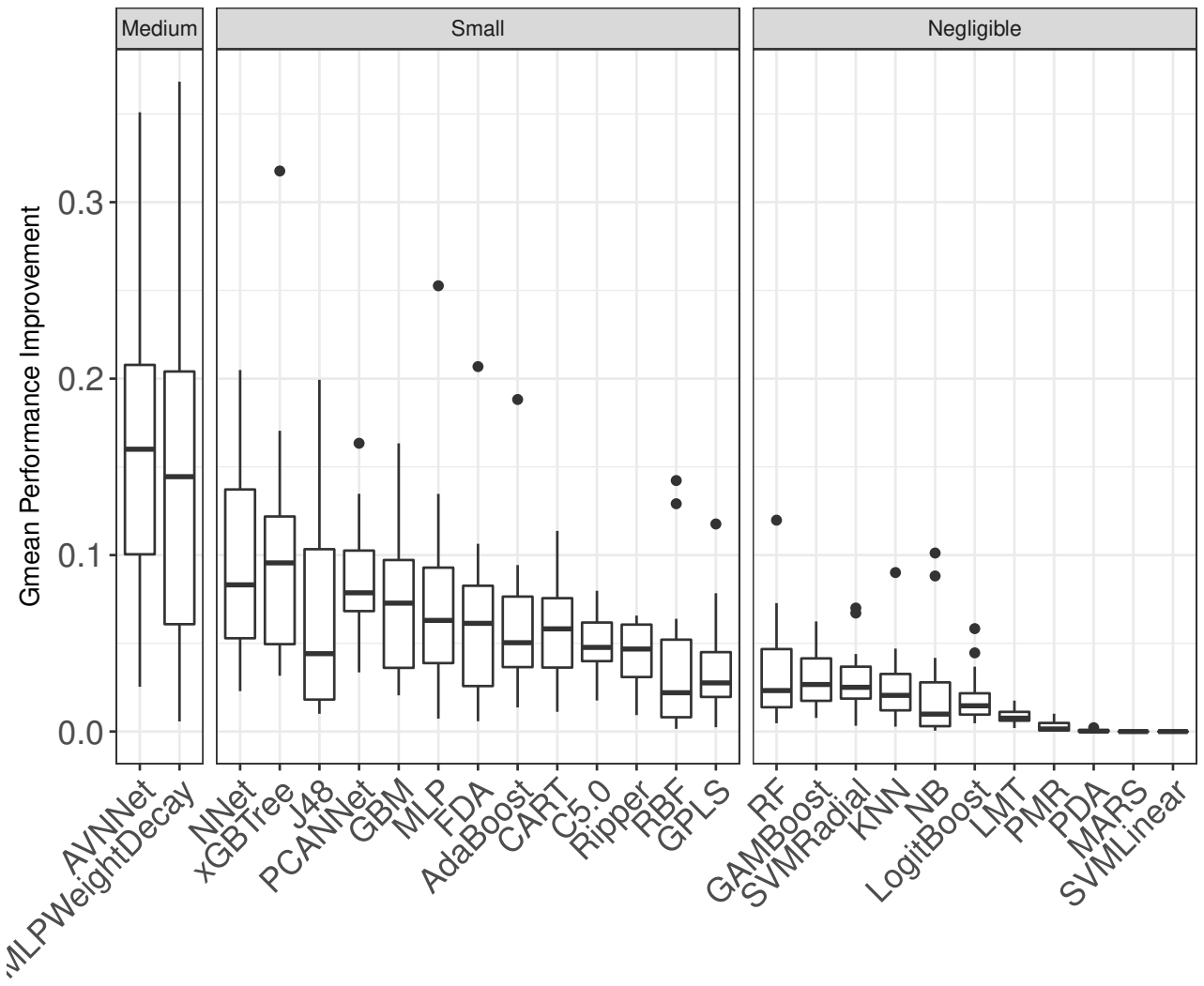

(g) Gmean Performance Improvement

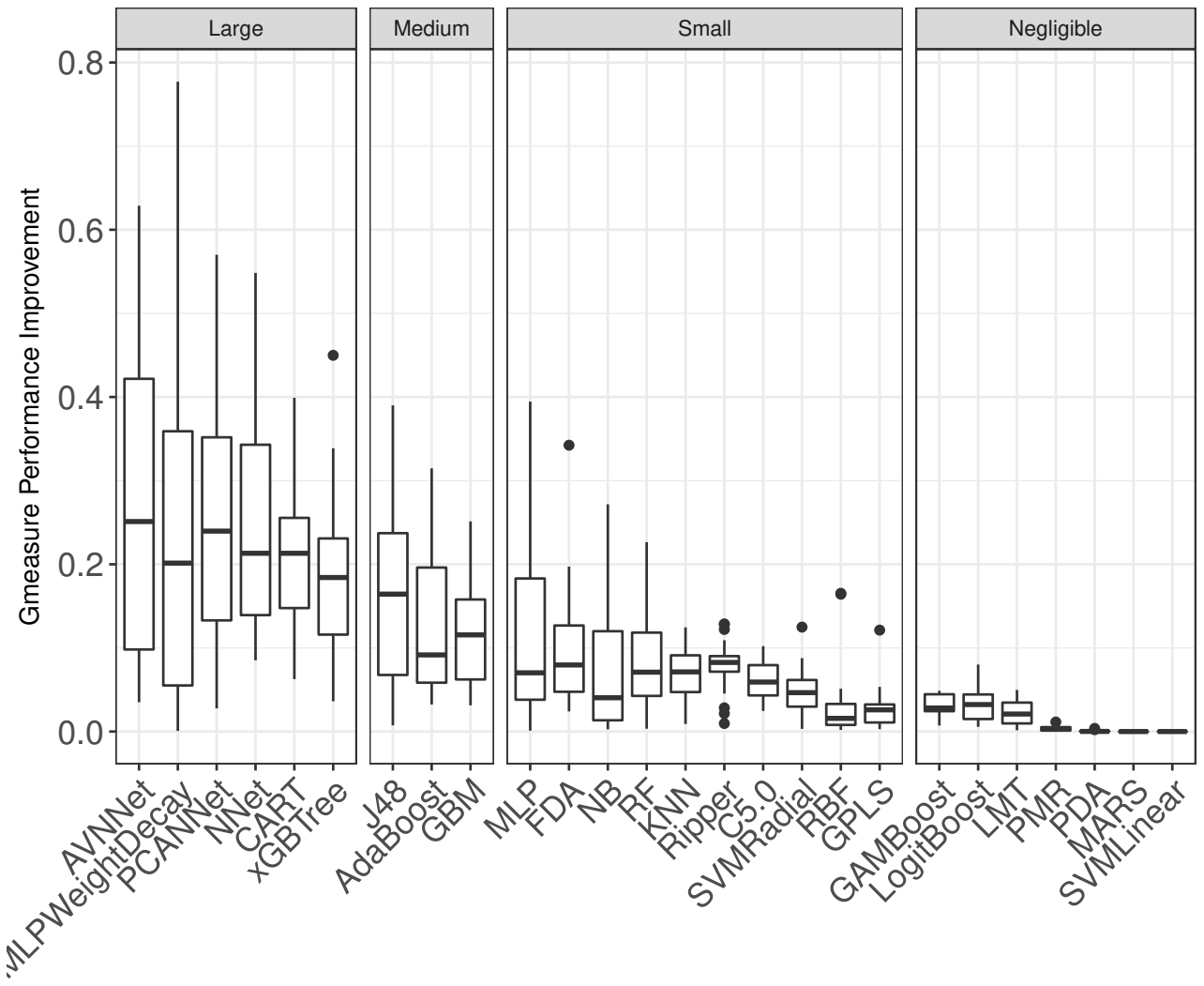

(h) Gmeasure Performance Improvement

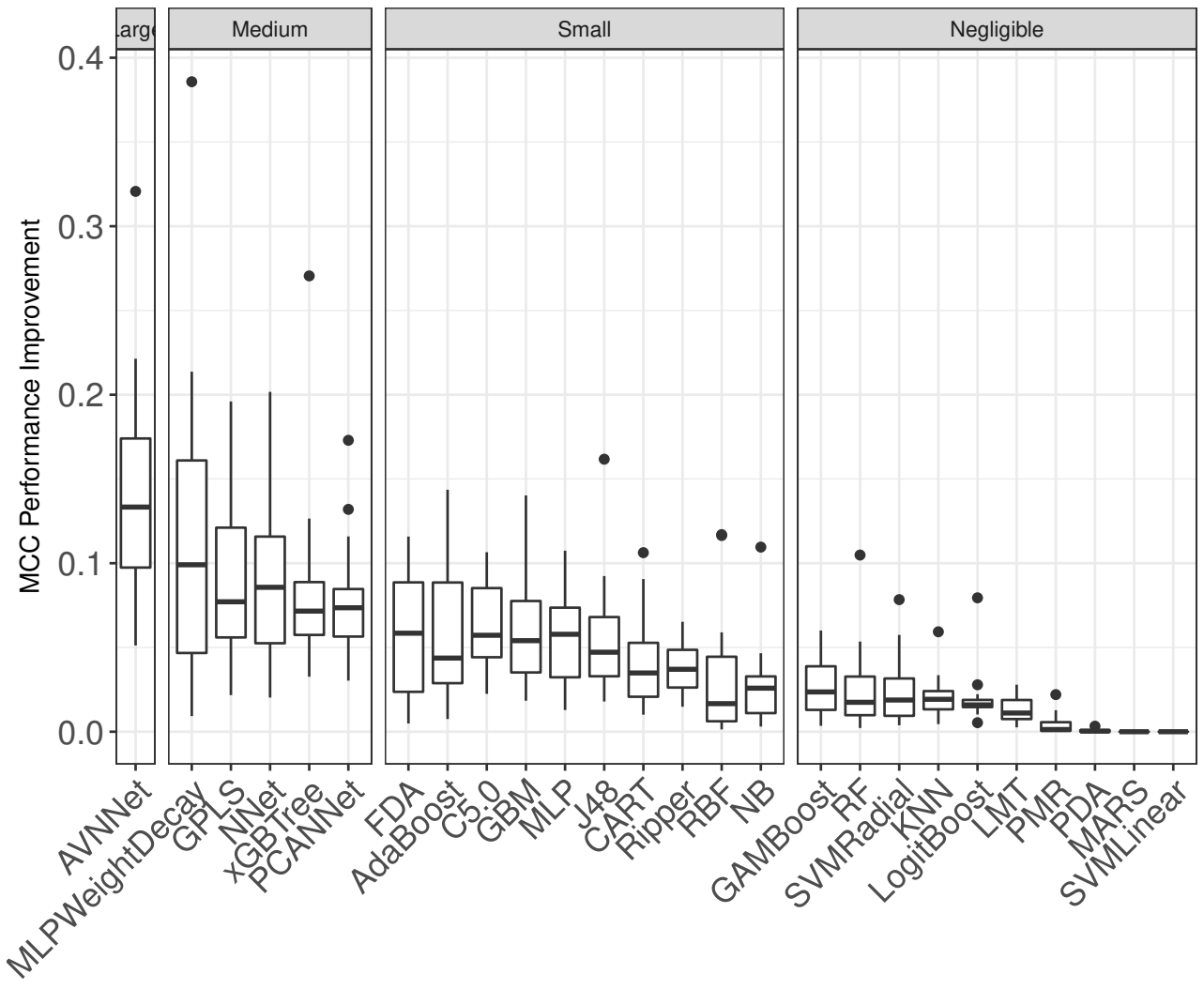

(i) MCC Performance Improvement

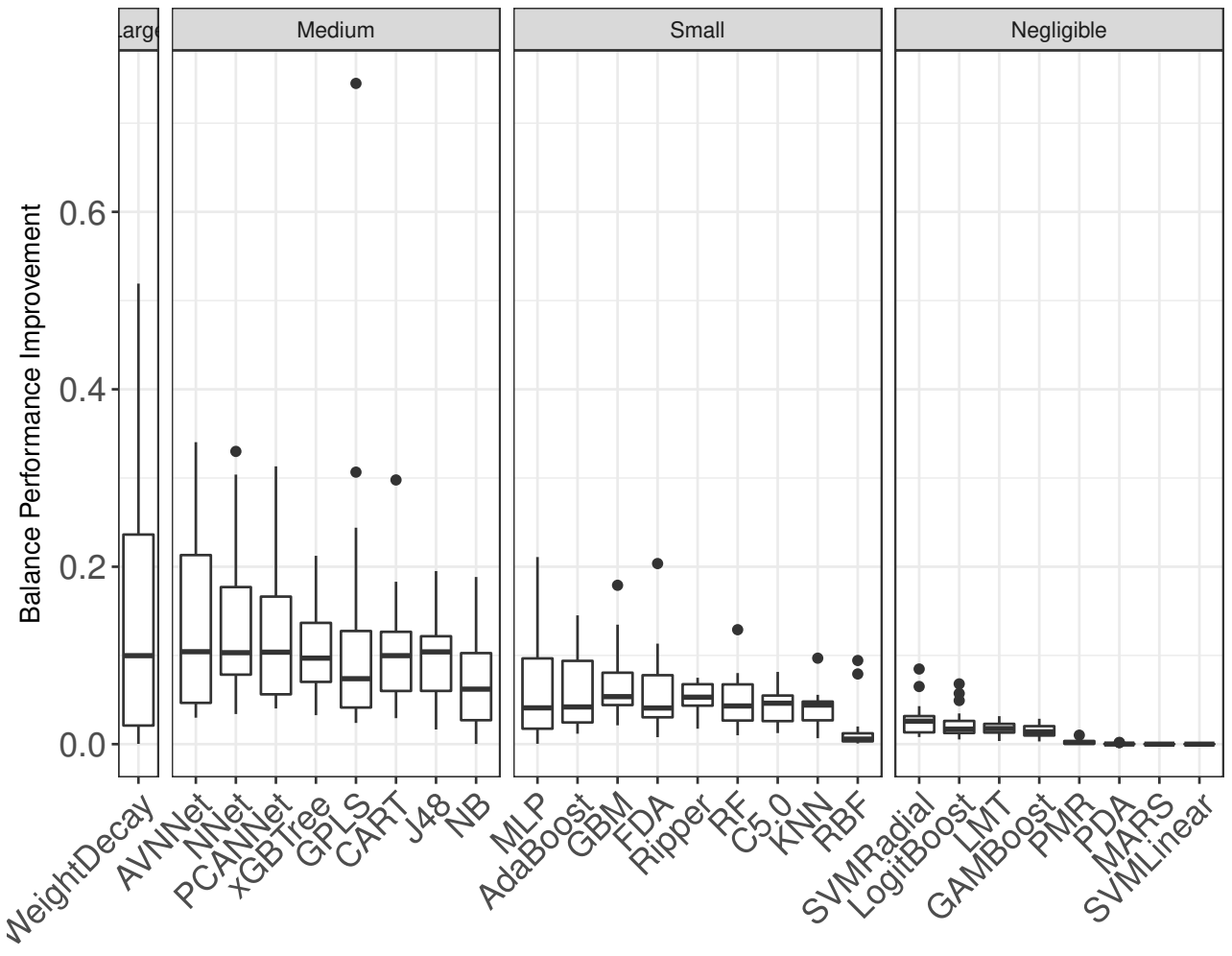

(j) Balance Performance Improvement

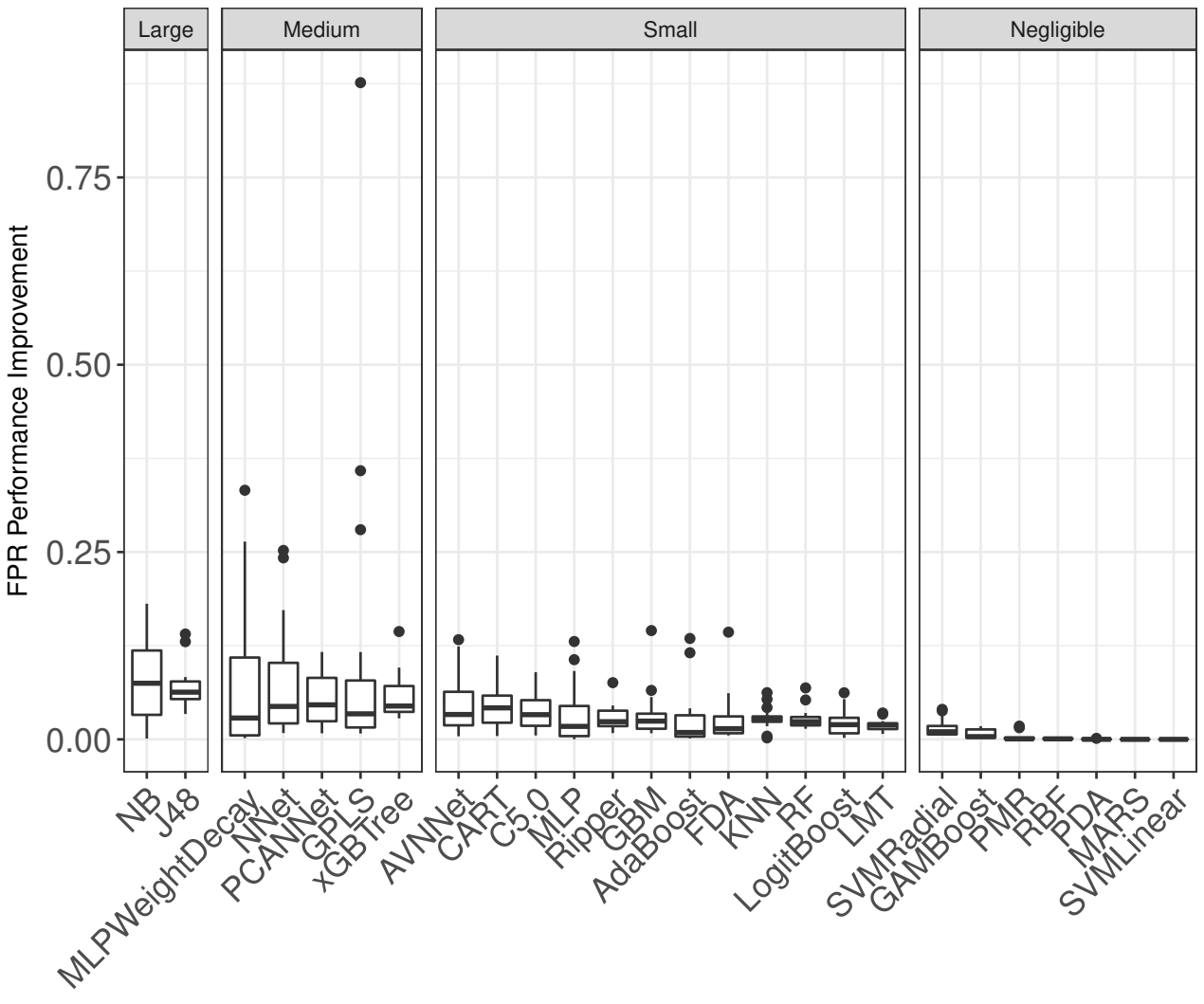

(k) FPR Performance Improvement

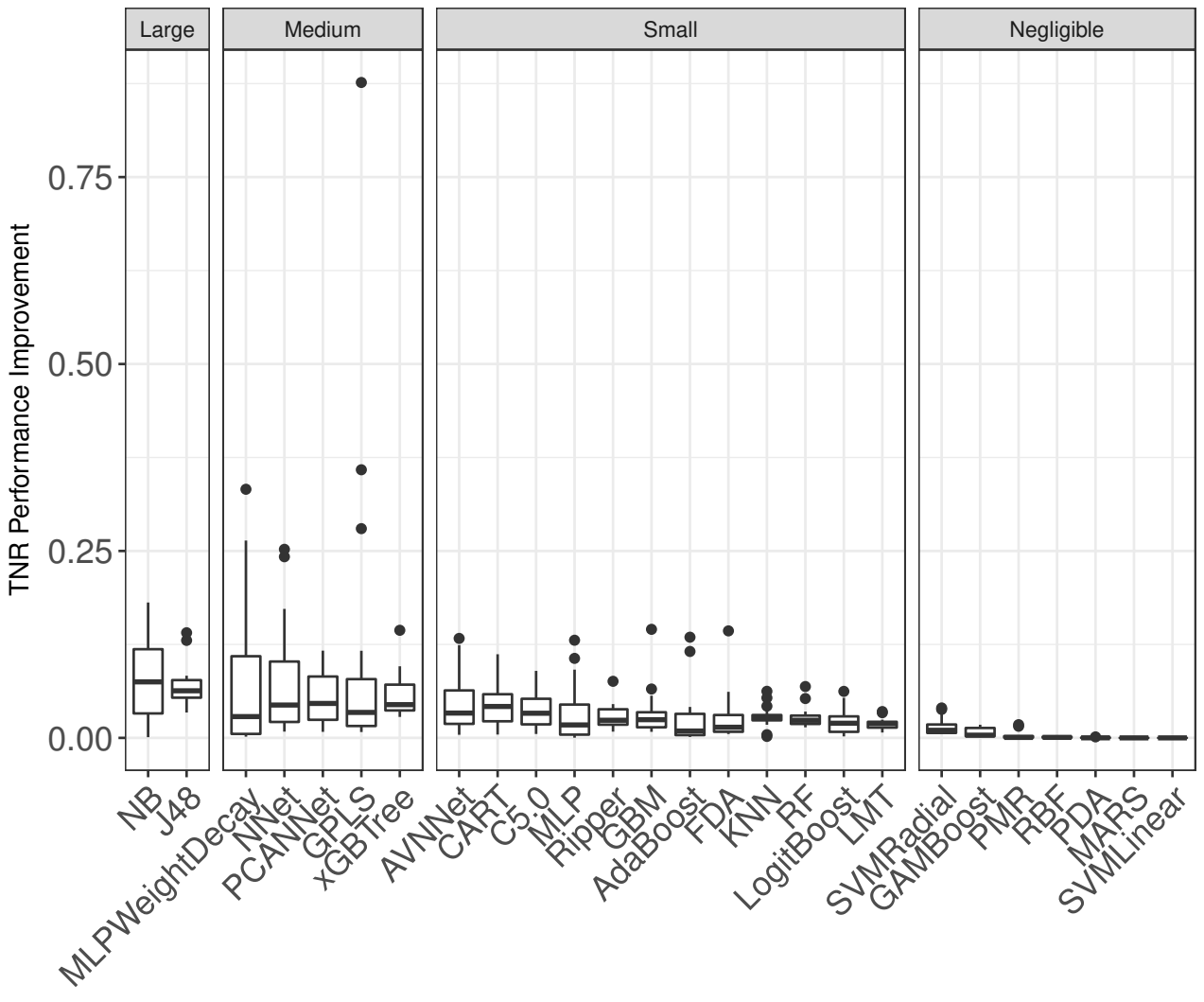

(l) TNR Performance Improvement

Fig. 4: (Continued) The performance improvement and its Cohen's $d$ effect size for each of the studied classification techniques.

the magnitude of the performance improvement, we use Cohen's $d$ effect size [21].

Furthermore, understanding the most influential parameters would allow researchers to focus their optimization effort. To this end, we investigate the performance difference for each of the studied parameters. To quantify the individual impact of each parameter, we train a classifier with all of the studied parameters set to their default settings, except for the parameter whose impact we want to measure, which is set to its optimized setting. We estimate the impact of each parameter using the difference of its performance with respect to a classifier that is trained entirely using the default parameter settings.

**Results**. **Optimization improves AUC performance by up to 40 percentage points.** Figure 4 shows the performance improvement for each of the 18 studied datasets and for each of the classification techniques. The boxplots show that optimization can improve the AUC performance by up to 40 percentage points. Moreover, the performance improvement provided by applying optimization is non-negligible (i.e., $d > 0.2$) for 16 of the 26 studied classification techniques (62%). This result indicates that parameter settings can substantially influence the performance of defect prediction models.

**Fortunately, random forest that is popularly-used in defect prediction studies tends to have negligible to small impact on the AUC performance of defect prediction models.** Figure 4 shows that optimization improves the performance of random forest classifiers as little as 2% of Brier and 5% of AUC. However, the random forest classifier tends to have a larger impact on threshold-dependent performance measures. Indeed, we observe that optimization improves the performance of random forest classifiers by up to 12% of Precision, 18% of Recall, and 15% of F-measure. This finding provides supporting evidence of Fu *et al.* [35] that automated parameter optimization impacts the threshold-dependent measures of random forest classifiers. Moreover, this finding suggests that random forest classifiers tend to robust to parameter settings when considering threshold-independent measures.

**C5.0 boosting yields the largest performance improvement when optimization is applied.** According to Cohen's $d$, the performance improvement provided by applying optimization is large for 9 of the 26 studied classification techniques (35%). On average, Figure 4 shows that the C5.0 boosting classification technique benefits the most from applying optimization, with a median performance improvement of 27 percentage points of AUC. Indeed, the C5.0 boosting classification technique improves from 6-40 percentage points of AUC. Moreover, irrespective to the choice of performance measures, `C5.0` classifier has non-negligible magnitude of performance improvement when optimization is applied.

Figure 5 shows that the `#boosting iterations` parameter of the C5.0 classification technique is the most influential parameter, while the `winnow` and `model type` parameters tend to have less of an impact. Indeed, the default `#boosting iterations` setting that is provided by the `C5.0` R package [77] is 1, indicating that only one C5.0 tree model is used for prediction. Nevertheless, we find that the optimal `#boosting iterations` parameter is 40, suggesting that the default parameter settings of the commonly-used research toolkits in literature nowadays are suboptimal for defect prediction datasets. This finding provides supporting evidence of the suspicions of prior studies [45, 63, 128, 146].

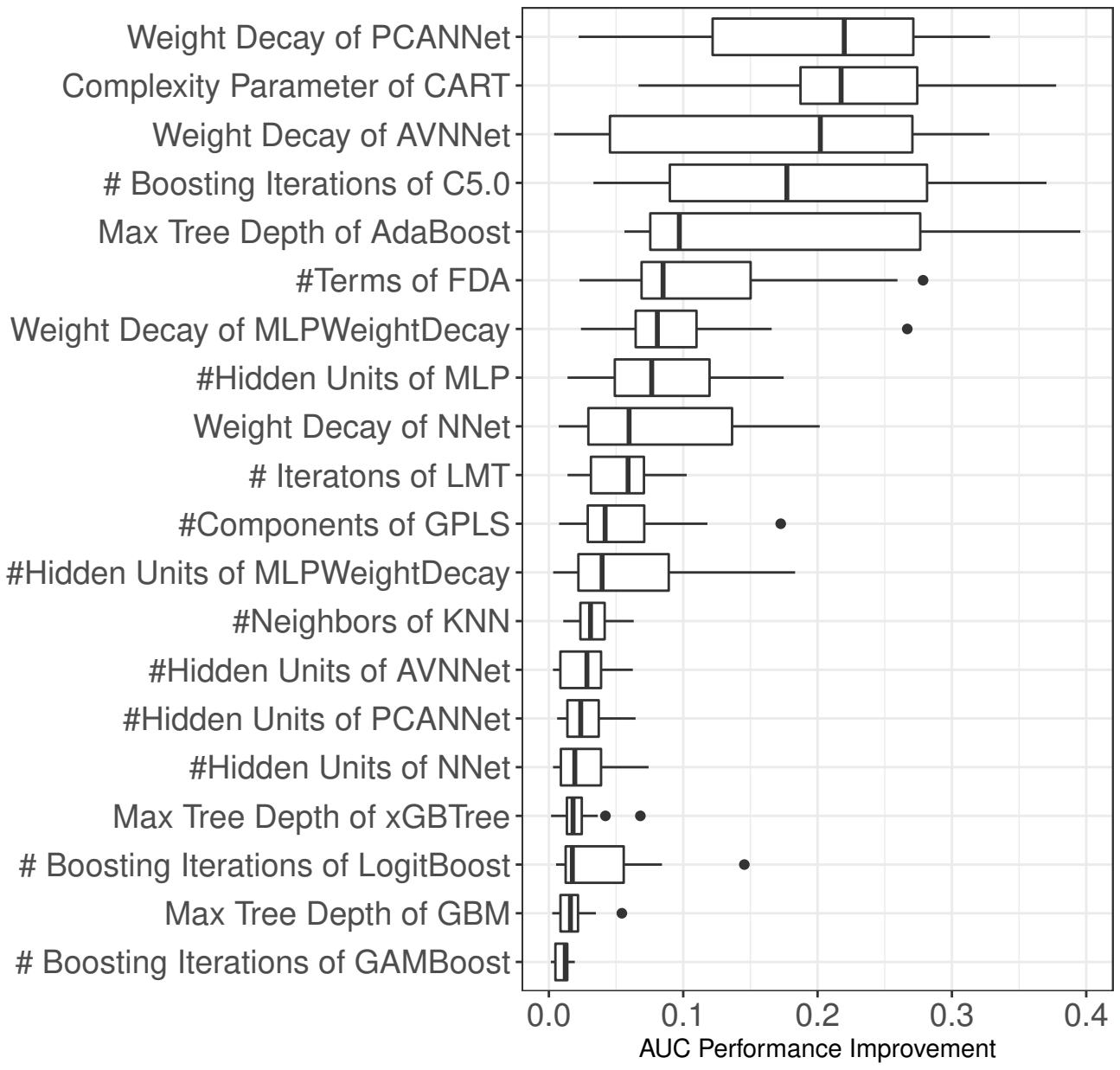

Fig. 5: The AUC performance improvement of the top-20 most sensitive parameters.

**In addition to C5.0 boosting, other classifiers like GPLS, and neural networks family (i.e., AVNNet, NNet, and PCANNet) also yield a considerably large benefit.** Figure 4 shows that the performance of the adaptive boosting (i.e., AdaBoost), advanced neural networks (i.e., AVNNet, PCANNet, NNet, MLP, and MLPWeightDecay), CART, and Flexible Discriminant Analysis (FDA) classification techniques have a median performance improvement from 13-24 percentage points with a large effect size. Indeed, Figure 5 shows that the fluctuation of the performance of the advanced neural network techniques is largely caused by changing the `weight decay`, but not the `#hidden units` or `bagging` parameters. Moreover, the `complexity` parameter of CART and `max tree depth` of adaptive boosting classification techniques are also sensitive to parameter optimization, indicating that researchers should pay more attention on the influential parameters and not to be as concerned on the less-influential one.

*Optimization improves the AUC performance of defect prediction models by up to 40 percentage points (e.g., C5.0 classifier). The performance improvement when applying optimization is non-negligible for 16 of the 26 studied classification techniques (62%), for example, C5.0, neural networks, and CART classifiers. Fortunately, the optimization of the parameters of random forest classifiers, one of the most popularly-used techniques in defect prediction studies, tends to have a negligible to small impact on the performance of random forest. Nevertheless, the impact of parameter optimization should be carefully considered when researchers select a particular technique for their studies.*

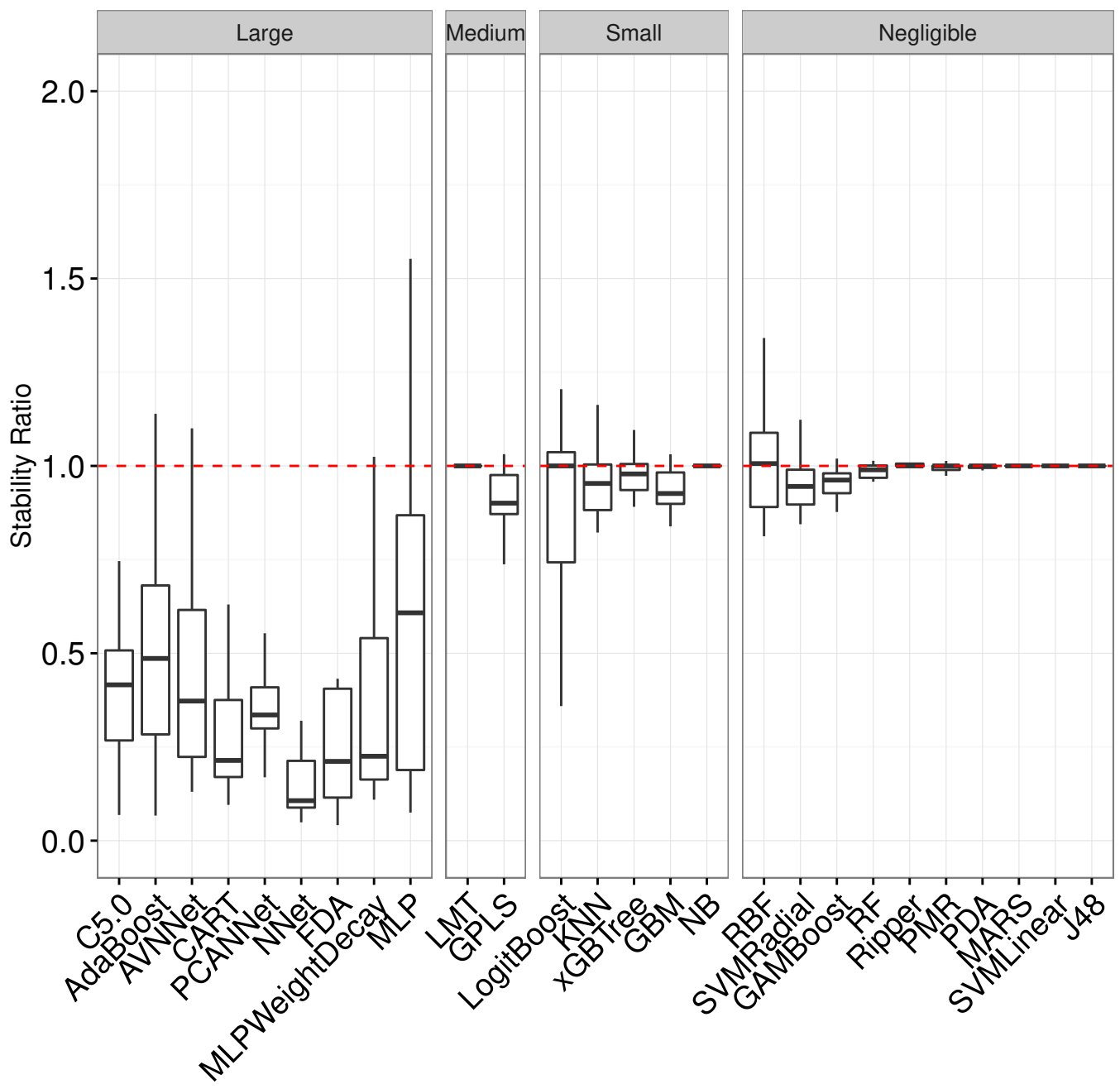

Fig. 6: The stability ratio of the classifiers that are trained using optimized settings compared to the classifiers that are trained using default settings for each of the studied classification techniques. A stability ratio that is less than one indicates that the classifiers that are trained with optimized settings are more stable than classifiers that are trained with default settings.

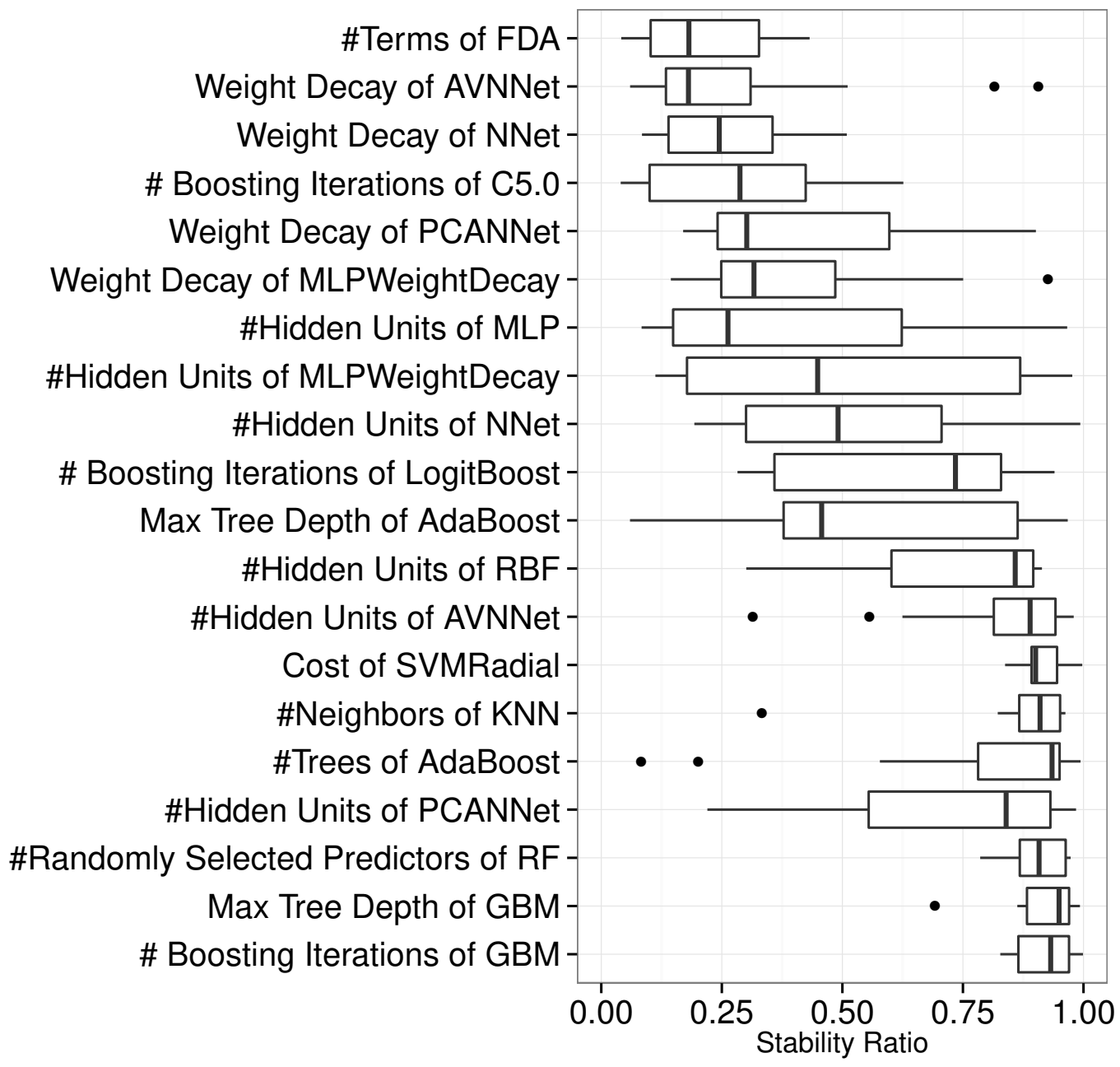

Fig. 7: The stability ratio of the top-20 most sensitive parameters.

*(RQ2) Does automated parameter optimization increase the performance instability of defect prediction models?*

**Approach**. To address RQ2, we start with the AUC performance distribution of the 26 studied classification techniques on each of the 18 studied datasets. The stability of a classification technique is measured in terms of the variability of the performance estimates that are produced by the 100 iterations of the out-of-sample bootstrap. For each classification technique, we compute the standard deviation (S.D.) of the bootstrap performance estimates of the classifiers where optimized settings are used ($\sigma_{optimized}$) and the S.D. of the bootstrap performance estimates of the classifiers where the default settings are used ($\sigma_{default}$). To analyze the difference of the stability between two classification techniques, we present the distribution of the stability ratio ($SR = \frac{\sigma_{optimized}}{\sigma_{default}}$) of the two classifiers when applied to the 18 studied datasets. A stability ratio that is less than one indicates that the classifiers that are trained with optimized settings are more stable than classifiers that are trained with default settings.

Similar to RQ1, we analyze the parameters that have the largest impact on the stability of the performance estimates. To this end, we investigate the stability ratio for each of the studied parameters. To quantify the individual impact of each parameter on model stability, we train a classifier with all of the studied parameters set to their default settings, except for the parameter whose impact we wish to measure, which is set to its optimized setting. We estimate the impact of each parameter using the stability ratio of its S.D. of performance estimates with respect to a classifier that is trained entirely using default settings.

**Results**. **Optimized classifiers are at least as stable as classifiers that are trained using the default settings.** Figure 6 shows that there is a median stability ratio of at least one for all of the studied classification techniques. Indeed, we find that the median ratio of one tends to appear for the classification techniques that yield negligible performance improvements in RQ1. These tight stability ratio ranges that are centered around one indicate that the stability of classifiers is not typically impacted by the optimized settings.

**The optimized classifiers of 9 of the 26 studied classification techniques (35%) are more stable than classifiers that are trained using the default settings.** Indeed, Figure 6 shows that there is a median stability ratio of 0.11 (NNet) to 0.61 (MLP) among the 9 classification techniques where the stability has improved. This equates to a 39%-89% stability improvement for these optimized classifiers. Indeed, Figure 7 shows that the stability of the performance of the advanced neural network techniques is largely caused by changing the `weight decay`, but not the `#hidden units` or `bagging` parameters, which is consistent with our findings in RQ1.

*Optimized classifiers are at least as stable as classifiers that are trained using the default settings. Moreover, the optimized classifiers of 9 (including C5.0, neural networks, and CART classifiers) of the 26 studied classification techniques (35%) are significantly more stable than classifiers that are trained using the default settings. On the other hand, for random forest and naïve bayes classifiers, the performance of the default-settings models is as stable as the optimized models.*

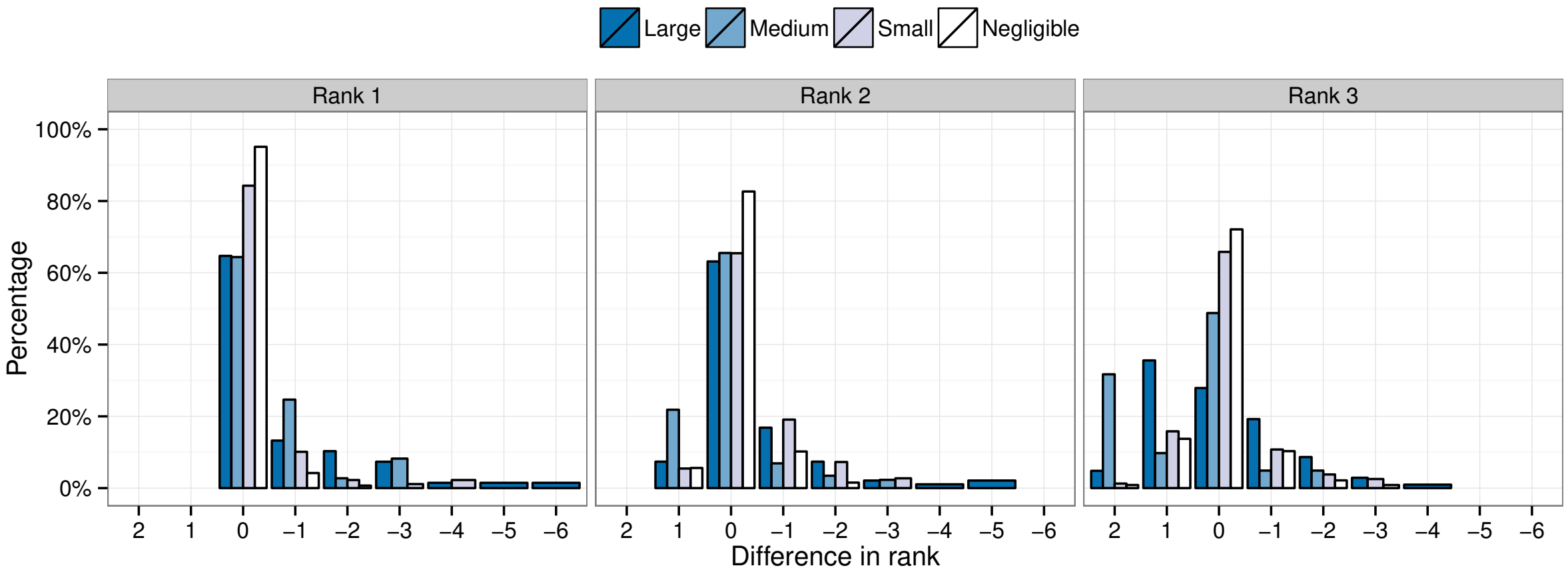

Fig. 8: The difference in the ranks for the variables according to their variable importance scores among the classifiers that are trained using optimized settings and classifiers that are trained using default settings. The bars indicate the percentage of variables that appear at that rank for the optimized model while also appearing at that rank for the default-setting models. The colours of the bars indicate the effect size of the performance improvement of classifiers according to Figure 4a.

*(RQ3) How much does the interpretation of defect prediction models change when automated parameter optimization is applied?*

**Approach**. To address RQ3, we start with the variable ranking of the 26 studied classification techniques on each of the 18 studied datasets for both optimized and default models. For each classification technique, we compute the difference in the ranks of the variables that appear in the top-three ranks of the classifiers that are trained using the optimized and default settings. For example, if a variable $v$ appears in the top rank in both the optimized and default-settings models, then the variable would have a rank difference of 0. However, if $v$ appears in the third rank in the default-settings model, then the rank difference of $v$ would be -2.

**Results**. **The insights that are derived from 42% of the optimized classifiers are different from the insights that are derived from the classifiers that are trained using the default settings.** Figure 8 shows the rank differences for all of the studied classifiers. In the 11 studied classification techniques that yield a medium to large performance improvement, we find that 35%-36% of the variables in the top importance rank of the optimized models do not appear in the top importance rank of the default-settings models. Indeed, in the 9 studied classification techniques that yield a large performance improvement, as few as 28% of the variables in the third rank of the optimized models also appear in the third rank of the default-settings models. This indicates that the interpretation of defect prediction models is heavily influenced by the parameter settings of classification techniques that are used to train the models.

**Classification techniques that are sensitive to parameter settings in the performance tend to also be sensitive in the interpretation.** Figure 8 shows that a smaller percentage of the variables of the optimized models are found at the same rank in the classification techniques that have a large performance improvement than the classification techniques that have a negligible performance improvement. Indeed, 65% of the variables in the top importance rank of classification techniques that yield a large performance improvement are found at the same rank. On the other hand, 95% of the variables appear at the same rank in the optimized and default-settings models of classification techniques where optimization yields a negligible performance improvement (e.g., random forest classifiers). This indicates that automated parameter optimization must be applied to classification techniques that are sensitive to parameter settings. Moreover, the interpretation of defect prediction studies that apply such sensitive classification should be revisited in order to confirm the stability of the findings.

*Optimization substantially shifts the importance ranking of variables, with as few as 28% of the top-ranked variables from the optimized models appearing in the same ranks as the non-optimized models. Classification techniques where optimization has a large impact on performance (e.g., C5.0, neural networks, CART classifiers) are subject to large shifts in interpretation. On the other hand, 95% of the variables appear at the same rank in the classification techniques where optimization has a negligible performance improvement (e.g., random forest classifiers).*

*(RQ4) How well do optimal parameter settings transfer from one dataset to another?*

**Approach**. For each classification technique, we obtain a list of optimal settings for every studied dataset. We then estimate the transferability of these settings using the frequency of their appearance in the Caret suggestions across datasets. We refer to settings that appear in the Caret suggestions of several datasets as "transferable".

We analyze 4 different types of transferability, i.e., across the 18 datasets, across the 5 proprietary datasets, across the 3 Eclipse datasets, and across the 2 NASA datasets. Since the transferability of optimized parameters has more of an impact on the most sensitive parameters, we focus our analysis on the 20 most sensitive parameters (see Figure 5).

**Results**. **The optimal setting of 17 of the 20 most sensitive parameters can be applied to datasets that share a similar set of metrics without a statistically significant drop in**

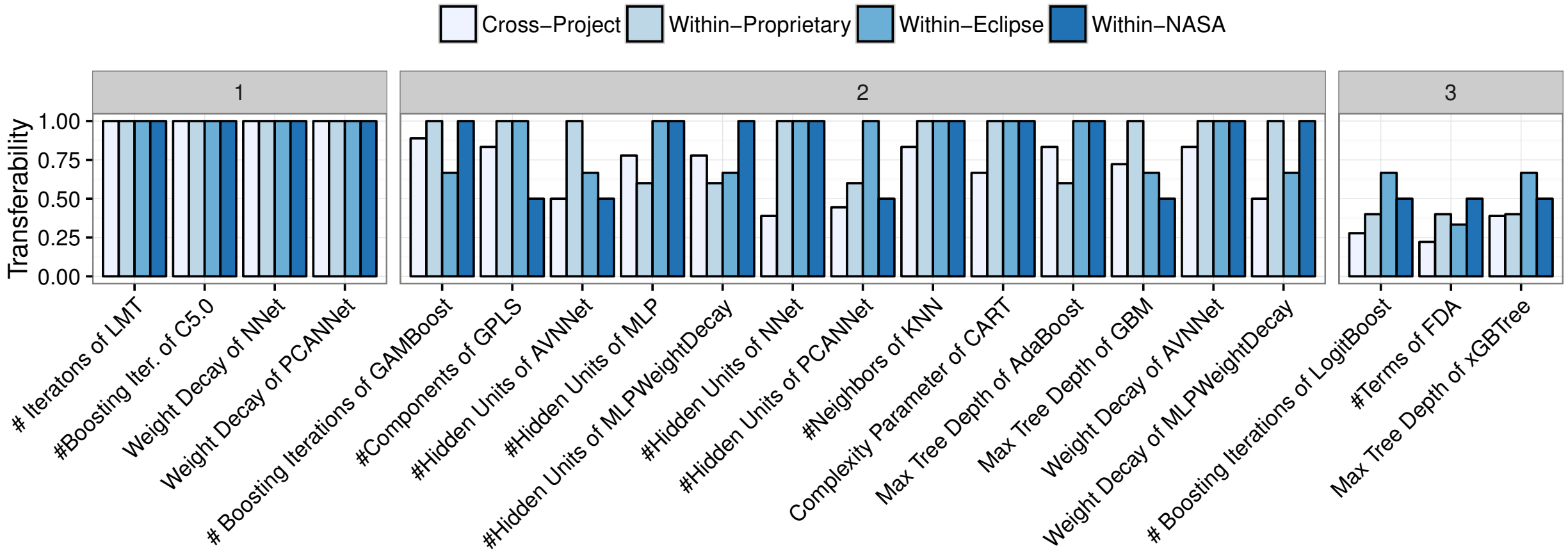

Fig. 9: Four types of transferability for each of the top-20 most sensitive parameters. A higher frequency for each of the optimized settings that appear across datasets indicates high transferability of a parameter.

**performance.** Figure 9 shows the 4 types of transferability for each of the 20 most sensitive parameters. We divide the transferable parameters into three groups: (1) those that can be transferred across the studied datasets, (2) those that can be transferred across some of the studied datasets, and (3) those that cannot be transferred across any of the studied datasets.

We find that the optimal setting of 17 of the 20 most sensitive parameters can be applied to other datasets that share a similar set of metrics without a statistically significant drop in performance with respect to the optimal performance for that dataset (i.e., the first and the second groups). For example, the value 40 for the `# boosting iterations` parameter of the C5.0 boosting classification technique can be applied to all of the 18 studied datasets without a statistically significant drop in performance. Moreover, the optimal value of 9 for `#hidden units` and value of 0.1 for `weight decay` parameters of the advanced neural networks (i.e., AVNNet) can be applied to 9 and 15 of the 18 studied datasets, respectively, that have a similar set of metrics without a statistically significant drop in performance. Indeed, the parameters of C5.0 and LMT are always transferable across the studied datasets, indicating that researchers and practitioners can safely adopt the optimized parameters that are obtained using a dataset with similar metrics. On the other hand, we find that only the parameters of LogitBoost, FDA, and xGBTree cannot be transferred across any of the studied datasets, indicating that researchers and practitioners should re-apply automated parameter optimization.

> *The optimal settings of 17 of the 20 most sensitive parameters can be applied to datasets that share a similar set of metrics without a statistically significant drop in performance. On the other hand, we find that only the parameters of LogitBoost, FDA, and xGBTree cannot be transferred across any of the studied datasets, indicating that researchers and practitioners should re-apply automated parameter optimization for such classifiers. However, we note that classifiers with low transferability are not top-performing classifiers.*

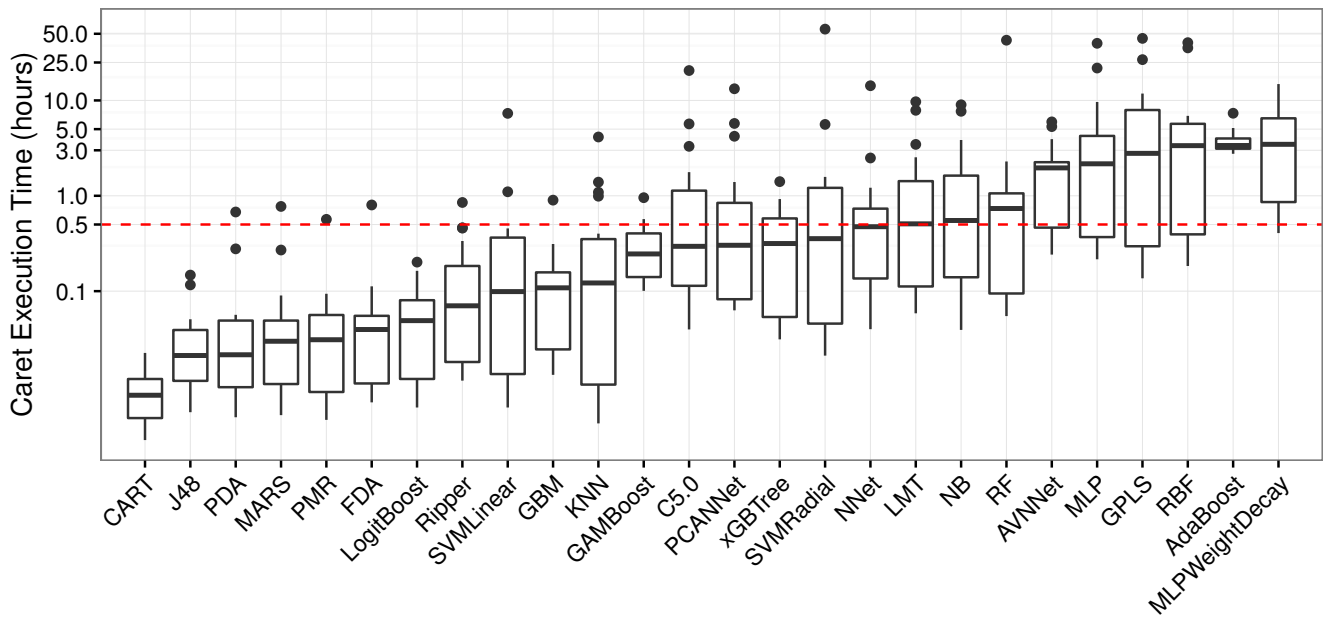

Fig. 10: The distribution of computational cost of grid-search optimization techniques (hours). The red-dashed line indicates the computational cost of 30 minutes.

*(RQ5) What is the computational cost of applying automated parameter optimization?*

**Approach**. Our case study approach is computationally-intensive (i.e., 450 parameter settings $\times$ 100 out-of-sample bootstrap repetitions $\times$ 18 systems $=$ 810,000 results). However, the results can be computed in parallel. Hence, we design our experiment using a High Performance Computing (HPC) environment. Our experiments are performed on 43 high performance computing machines with Intel Xeon 5675 @3.1 GHz (24 hyper-threads) and 64 GB memory.

For each classification technique, we compute the average amount of execution time that was consumed by grid-search optimization when producing the suggested parameter settings for each of the studied datasets.

**Results**. **Grid-search optimization adds less than 30 minutes of additional computation time to 12 of the 26 studied classification techniques.** The optimization cost of 12 of the 26 studied classification techniques (46%) is less than 30 minutes for at least 75% of the studied datasets. C5.0 and extreme gradient boosting classification techniques, which yield top-performing classifiers more frequently than other classification techniques, add less than 30 minutes of ad-ditional computation time for at least 50% of the studied datasets. This indicates that applying optimization tends to improve the performance of defect models while incurring a manageable additional computational cost.

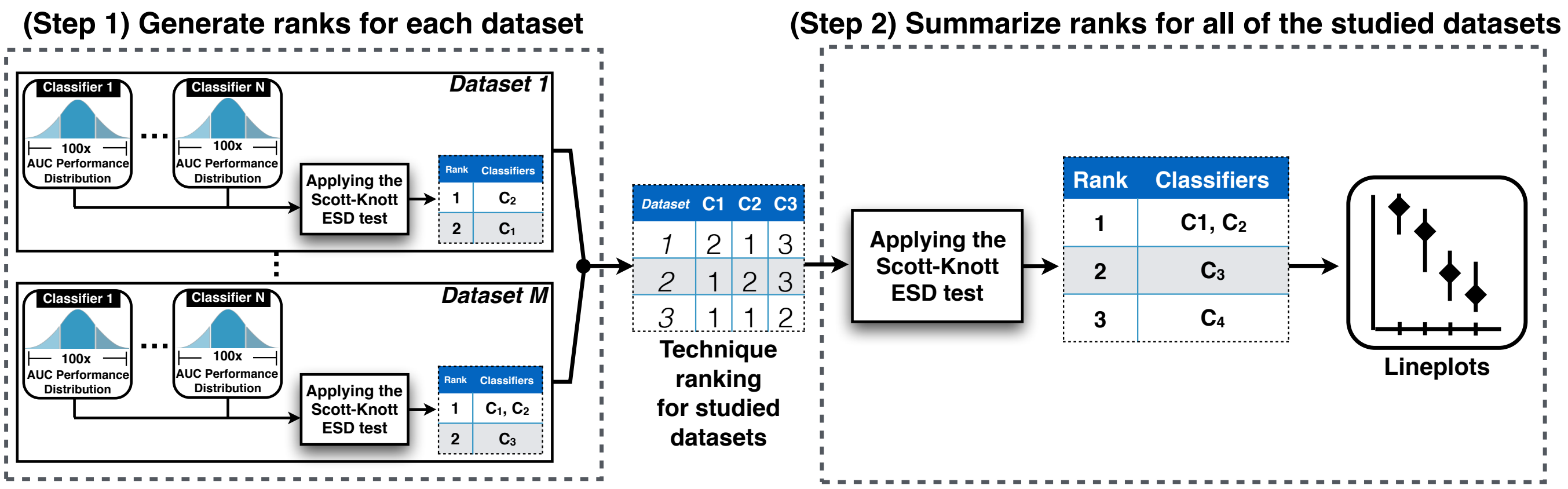

Fig. 11: An overview of the double Scott-Knott ESD approach (i.e., a statistical comparison over multiple datasets).

Moreover, the optimized classifiers of 9 of the 26 studied classification techniques (35%) (i.e., AdaBoost, MLPWeight-Decay, and RBF) require more than 3 additional hours of computation time to apply grid-search optimization for at least 50% of the studied datasets. We observe that the longer computational time of the other 50% studied datasets have to do with the size of the datasets. Indeed, the datasets that take the longest computational time are the top-3 largest studied datasets (i.e., `prop-1`, `prop-2`, and `eclipse-3.0`). Nonetheless, the computation could still be completed if it was executed overnight. Assuming that we used Amazon's EC2 cloud computing infrastructure to offload the optimization process, a `c4.large` container that provides 8 CPU and 3.75 GB of memory costs \$0.1 per hour (as of July 2017).[5] Therefore, the additional computational cost of the 3 slowest classification techniques can be offloaded to an Amazon EC2 instance for less than \$1. This back-of-the-envelope cost estimation suggests that the benefits of automated parameter optimization outweigh the costs. Since defect prediction models do not need to be (re)trained very often in practice, this cost should still be manageable.

> *Grid-search optimization adds less than 30 minutes of additional computation time to 12 of the 26 studied classification techniques. To assign an actual monetary cost for such optimization techniques, we used an Amazon EC2 costing estimates. We find that an Amazon EC2's estimated computational cost of the 3 slowest classification techniques is less than \$1.*

## 7 Revisiting the Ranking of Classification Techniques for Defect Prediction Models

Prior studies have ranked classification techniques according to their performance on defect prediction datasets. For example, Lessmann *et al.* [81] demonstrate that 17 of 22 studied classification techniques are statistically indistinguishable. On the other hand, Ghotra *et al.* [39] argue that classification techniques can have a large impact on the performance of defect prediction models.

Our earlier work [139] shows that automated parameter optimization increases the likelihood of appearing in the top Scott-Knott ESD rank by as much as 83%. However, our earlier work did not discuss the ranking of optimized and default-settings classifiers for defect prediction models. To address this, we formulate the following research question:

> *(RQ6) What is the ranking of optimized and default-settings classifiers for defect prediction models?*

### 7.1 Approach

To analyze the ranking of optimized and default-settings classifiers, we perform a double Scott-Knott ESD test [39]. The double Scott-Knott ESD approach is a statistical comparison approach over multiple datasets that consider dataset-specific performance. Figure 11 provides an overview of the double Scott-Knott ESD approach. The double Scott-Knott ESD test is divided into two steps that we describe below.

**(Step 1) Generate Scott-Knott ESD ranks for each dataset** Similar to Ghotra *et al.* [39], we first start with the AUC performance distribution of the 26 studied classification techniques that are trained with the optimized parameter settings and the default settings. To find statistically distinct ranks of classification techniques within each dataset, we provide the AUC performance distribution of the 100 bootstrap iterations of each classification technique with both parameter settings to a Scott-Knott ESD test ($\alpha = 0.05$) [141]. We use the Scott-Knott ESD test in order to control for dataset-specific model performance, since some datasets may have a tendency to produce over- or under-performing classifiers. Finally, for each classification technique, we have 18 different Scott-Knott ESD ranks (i.e., one from each dataset).

**(Step 2) Summarize ranks of classification techniques for all of the studied datasets.** We provide the 18 different Scott-Knott ESD ranks to a second run of the Scott-Knott ESD test to produces a statistically distinct ranks of classification techniques across all of the studied datasets.

5. https://aws.amazon.com/ec2/pricing/

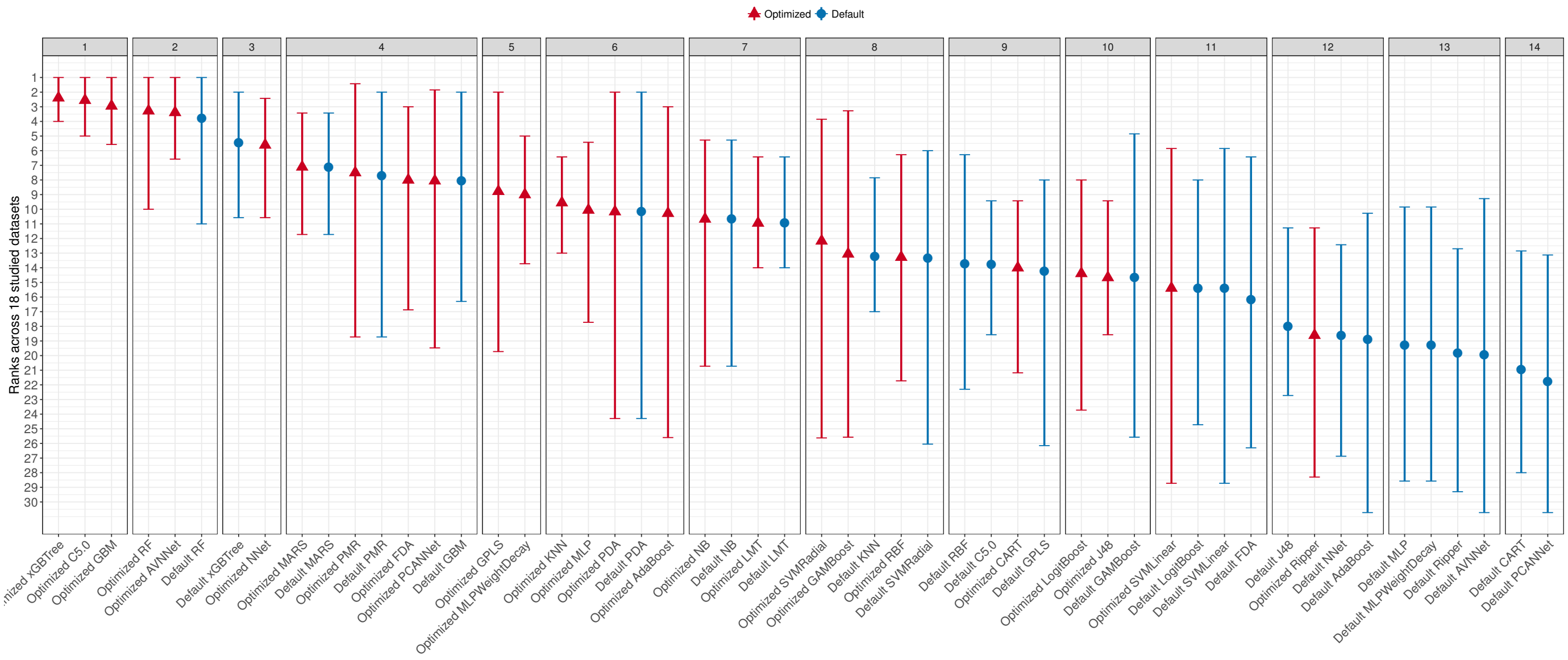

Fig. 12: The ScottKnott ESD ranking of optimized and default-settings classifiers across all of the studied datasets. Circle dots and triangle dots indicate the average ranks for optimized and default-settings models, respectively, while the lines indicate the 95% confidence interval of the ranks across the 18 studied datasets.

## 7.2 Results

Figure 16 shows the ScottKnott ESD ranking of optimized and default-settings classifiers across all of the studied datasets.

**Random forest is not the top-performing classification technique for all of the studied datasets.** Figure 16 shows that random forest, which is one of the most commonly-used classification technique in software engineering, appears at the second Scott-Knott ESD ranks. Unlike prior work in the data mining domain [30], we find that random forest is not the most frequent top performer in our defect prediction datasets. We find that the average rank of random forest is 3.277, where random forest appears at the top Scott-Knott ESD rank for 50% of the datasets. This contradicts the conclusions of Fernandez-Delgado *et al.* [30], who find that random forest tends to yield top-performing classifiers the most frequently on non-software engineering datasets. The contradictory conclusions suggest that domain-specifics play an important role.

Nevertheless, we find that optimization does not impact the ranking of random forest. Figure 16 shows that both optimized and default-settings random forest classifiers appear at the second Scott-Knott ESD ranks. This result echoes the findings of RQ1, where we noted that random forest yileds the negligible to small benefits of applying automated parameter optimization techniques.

**Only the optimized extreme gradient boosting trees (xGBTree), C5.0, gradient boosting machine (GBM) classifiers are the top-performing classification techniques across all of the studied datasets.** Figure 16 shows that there are only three optimized classifiers (i.e., xGBTree, C5.0, and GBM) are the top-performing classification techniques across all of the studied datasets. We find that the average ranks of optimized xGBTree, C5.0, and GBM classifiers are 2.39, 2.56, and 2.94, respectively. Indeed, we find that automated parameter optimization substantially changes the ranking of C5.0 classifiers from the ninth rank to the first rank. This result echoes the findings of RQ1, where we noted that C5.0 boosting benefits the most of automated parameter optimization.

**Automated parameter optimization substantially improves the ranking of default-settings neural networks, C5.0, gradient boosting machines, and CART classifiers.** We observed a trend that optimized classifiers (i.e., red lines) tend to appear at the top Scott-Knott ESD ranks (i.e., left hand) more than default-settings classifiers (i.e., blue lines). Figure 16 shows that 18 of the 26 studied classification techniques (70%) (e.g., C5.0, GBM, xGBTree, AVNnet, NNet, and CART) that are trained with optimized settings appear at the lower rank than that classification techniquest that are trained with default settings. Morever, we find that the optimized AVNNet classifiers and random forest classifiers appear at the same Scott-Knott ESD rank, suggesting that future studies on Deep Learning in software engineering should consider automated parameter optimization. This finding also echoes the importance of a study of Fu *et al.* [34] where automated parameter optimization is essential for neural network and deep learning in software engineering domain.

*Random forest is not the top-performing classification technique for all of the studied datasets, which disagrees with the findings of prior general data mining and recent software engineering related studies. In addition, some rarely-used low-performing classification techniques (e.g., C5.0) are substantially outperforming widely-used techniques like random forest—highlighting the importance of exploring the parameter settings of classifiers.*

# 8 An Empirical Comparison of Automated Paramter Optimization Techniques for Defect Prediction Models

Recent studies show several benefits of applying automated parameter optimization techniques. For example, Tantithamthavorn *et al.* [139] show the improvement of the performance and stability of defect prediction models when grid-search parameter optimization is applied. Fu *et al.* [35] show that the performance of defect prediction models can be improved when applying a differential evolution algorithm. Moreover, Di Martino *et al.* [27] and Sarro *et al.* [121] show that the performance of support vector machines is substantially improved when using a genetic algorithm as an automated parameter optimization. Hu *et al.* [57] adopt a genetic algorithm to find an optimal parameter setting of neural network for defect prediction models.

Although recent works reach the same conclusions that automated parameter optimization techniques should be included in future defect prediction studies, they apply different parameter optimization techniques. Indeed, the choice of parameter optimization techniques may impact the conclusions of recent defect prediction studies. For example, Tantithamthavorn *et al.* [135, 140] and Menzies *et al.* [97] point out that the use of different techniques for an experimental component may influence the conclusions of defect prediction studies. Yet, little research is known about the impact that parameter optimization techniques have on the performance of defect prediction models. Indeed, it is not clear which optimization techniques should be used for future defect prediction studies. The lack of consistency in the choice of parameter optimization techniques of prior work makes it hard to derive practical guidelines about the most appropriate parameter optimization techniques to use in future defect prediction research. To address this, we formulate the following research question:

***(RQ7)*** *Which optimization techniques yield the largest performance improvement for defect prediction models?*

## 8.1 The State-of-the-Art of Automated Parameter Optimization Techniques

There are a plethora of automated parameter optimization techniques that can be applied [9, 10, 79, 87]. Since it is impractical to study all of these techniques, we would like to select a manageable, yet representative set of optimization techniques for our study. To do so, we analyze the Search-Based Software Engineering (SBSE) literature in order to identify the commonly used search-based techniques.

We first start with a grid search technique that is previously used in our recent work [139]. Based on a recent literature review of Harman *et al.* [48], we add a random search technique [10], and two evolutionary algorithms that include a genetic algorithm [41] and a differential evolution algorithm [132]. An evolutionary-based search technique directs the search using an approach that resembles the biological evolution and natural selection processes. Below, we provide the description of the parameters for each automated parameter optimization technique.

### *8.1.1 Grid Search Technique*

Grid search technique is the simplest optimization technique based on an exhaustive search through a manually specified set of the parameter space of a classification technique [7, 8, 56]. The advantage of using grid search technique is that the candidate parameter settings are systematically generated. Thus, each dataset will be evaluated using the same candidate parameter settings.

<u>Parameter:</u>

– `Budget` indicates the number of different values to be evaluated for each parameter.

Grid search is made up of three steps as follows:

1) Systematically generate candidate parameter settings based on a given `budget` threshold. For example, a `budget` threshold of 5 will limit the number of candidate settings for each parameter to 5. Thus, for an arbitrary classifier with 3 parameters and a `budget` threshold of 5, grid search technique generates $5 * 5 * 5 = 125$ combinations of parameter settings.
2) Evaluate each candidate parameter setting.
3) Identify optimal parameter setting.

### *8.1.2 Random Search Technique*

Unlike grid search technique, random search technique exhaustively searches through a randomly generated set of the parameter space of a classification technique [112]. Bergstra and Bengio [10] argue that, given the same computational cost, advanced parameter optimization techniques (e.g., random search, genetic algorithm, differential evolution) can find a more accurate model due to a larger search space of candidate parameter settings.

<u>Parameter:</u>

– `Iteration` indicates maximum number of the combination of different parameter settings to be evaluated for a classification technique.

Random search is made up of three steps as follows:

1) Unlike grid search technique that needs pre-defined candidate parameter settings, a random search technique randomly generates candidate parameter settings based on a given `iteration` threshold. For example, an `iteration` threshold of 5 will limit the number of candidate parameter settings for each classification technique to 5. Thus, regardless of the number of parameter settings, the random search technique always generates 5 combinations of parameter settings for a classification technique.
2) Evaluate each candidate parameter setting.
3) Identify optimal parameter setting.

### *8.1.3 Genetic Algorithm (GA)*

A genetic algorithm is an evolutionary-based optimization technique which is based on natural selection and genetics concepts, where chromosomes of the two individuals (parents) work together to form new individual by keeping the best property from each of the parent individuals [41]. A genetic algorithm optimizes a population of parameter settings for a given objective, e.g., maximizing the performance of defect prediction models. In a genetic algorithm, two parents are selected from the population for reproduction (*crossover*), i.e., generating offspring. The generated offspring is a combination of parameter settings that of values are shared with at least one parent. To drive the evolution, a genetic algorithm randomly changes one or more parameter settings of the offspring (*mutation*). Finally, the performance of the mutated offspring is evaluated and either selected as an optimal combination of parameter setting or inserted into the population for reproduction in a next generation.
Parameters:

- `Population size` indicates the size of the population. This parameter defines how many offspring will be generated in parallel during a generation.
- `Crossover probability` indicates the probability of a parameter setting behing inherited from one of the parents.
- `Mutation probability` indicates the probability of a parameter setting mutating from the inherited value.
- *Elitism* indicates the number of best-performing combinations of parameter settings that are guaranteed a role as parent or as optimal combination in the next generation.
- *Generations* indicates the number of generations with no consecutive performance improvement before the algorithm will stop.

A genetic algorithm is made up of five steps as follows:

1) Initialize the Population. The population is initialized by randomly generating a combination of parameter settings given the `population size` threshold.
2) Evaluate each candidate parameter setting.
3) Randomly select pairs of parents from the population and generate a mutated offspring using the *crossover* and *mutation probabilities*. Mutation allows the algorithm to introduce diversity into the population, expanding the opportunity to search unexplored areas in the search space for fitter solutions.
4) Select the best-performing combinations into the population, taking *elitism* into account.
5) Return to step 2 until the algorithm terminates.

#### *8.1.4 Differential Evolution (DE)*

Differential evolution is an evolutionary-based optimization technique [132], which is based on differential equation concept. Unlike a GA technique that uses crossover as search mechanisms, a DE technique uses mutation as a search mechanism. Hegerty *et al.* [52] point out that a DE technique provides a more stable optimal solution than genetic algorithm. Moreover, Hegerty *et al.* [52] also point out that a DE technique finds better global solution than a genetic algorithm.
Parameters:

- `Population size` indicates the number of multiple candidate (individual) generated at each generation.
- `Crossover probability` is the probability of crossover taking place between two individuals for mutation process.
- `Strategy` determines the function for mutation in differential evolutionary technique.

Differential Evolution is made up of three steps as follows:

1) (Population initialization step): DE randomly generates the initial population of candidate parameter settings, given a `population size` threshold.
2) (Mutation step): DE generates new candidates by adding a weighted difference between two population members to a third member based on a `crossover probability` parameter.
3) (Selection step) Keep the best candidate parameter setting for the next generation.

### 8.2 Approach

We use the high-level approach of Figure 1 that is described in Section 5 to address RQ7. While Steps 1 (i.e., Generate Bootstrap Sample) and 4 (Calculate Performance) of the high-level approach is identical, Steps 2 (i.e., Identify Optimizal Setting) and 3 (Construct Defect Prediction Models) are performed differently. We describe the different Steps 2 and 3 below.

#### *8.2.1 (Step 2) Identify optimizal settings*

To study the impact of optimal settings that are derived from different automated parameter optimization techniques, we only focus on off-the-shelf parameter optimization techniques that are readily-available so our observations and recommendations are easy to transfer to industrial practices. Thus, we apply four automated parameter optimization techniques as discussed in Section 8.1 (i.e., Grid Search, Random Search, Genetic Algorithm, and Differential Evolution). We use the implementation of a grid search technique that is provided by the `train` function of the `caret` R package [78] by specifying the following argument `trainControl(search="grid")`. We use the implementation of a random search technique that is provided by the `train` function of the `caret` R package [78] by specifying the following argument `trainControl(search="random")`. We use the implementation of a genetic algorithm that is provided by the `ga` function of the `GA` R package [122]. We use the implementation of a differential evolution that is provided by the `DEoptim` function of the `DEoptim` R package [100].

#### *8.2.2 (Step 3) Construct defect prediction models*

In selecting the classification techniques for our empirical comparison, we identified three important criteria that need to be satisfied:

- **Criterian 1 – Configurable classification techniques:** Since we want to study the impact that optimization techniques have on the performance of defect prediction models, we select only the classification techniques that require at least one configuration parameter setting (see Section 2).
- **Criterian 2 – Sensitive classification techniques:** The results of RQ1 the sensitivity of the parameters of classification techniques for defect prediction models are different (see Section 6). Thus, we opt to study only the classification techniques whose parameter settings have a large impact on the performance of defect prediction models.

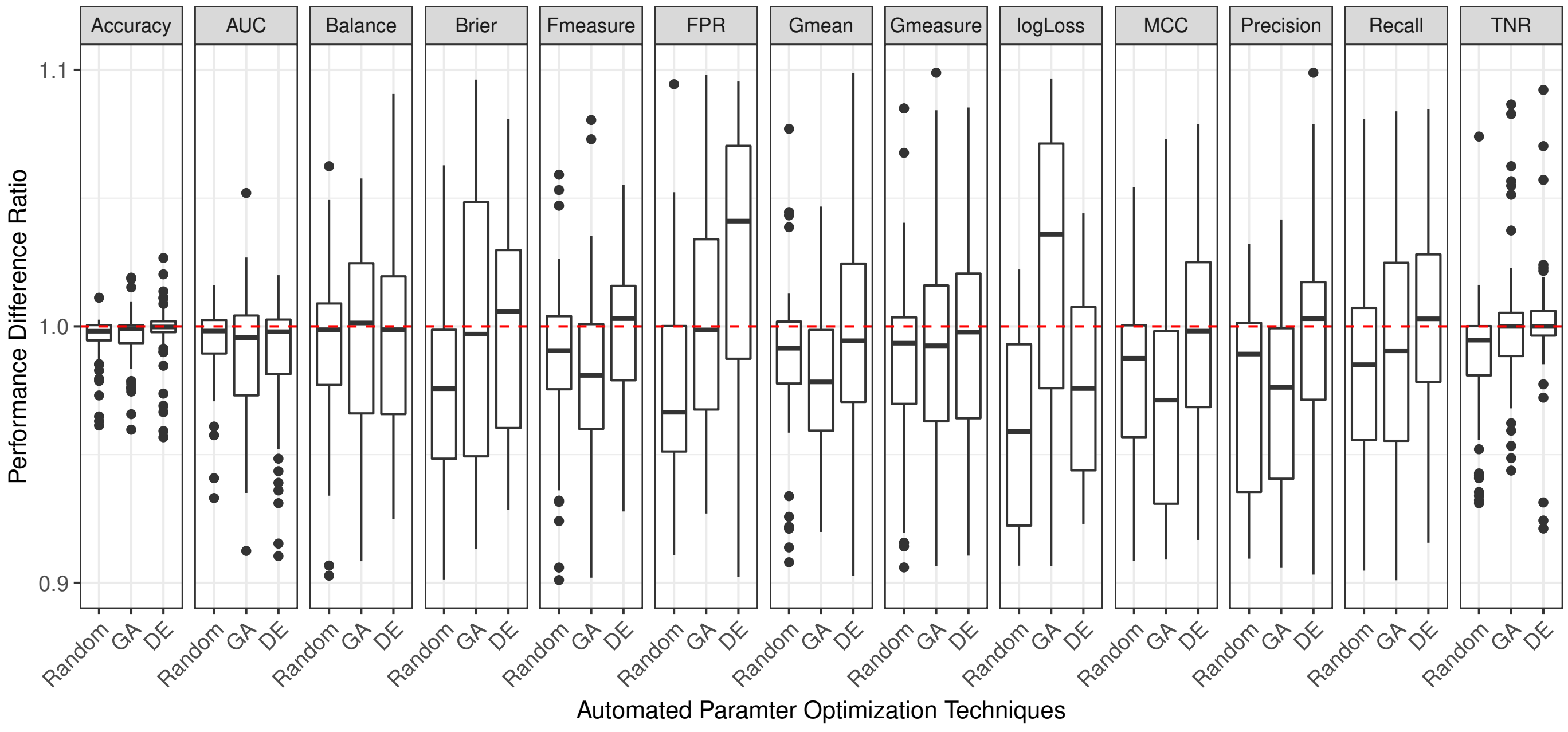

Fig. 13: The distributions of the performance difference ratio of each performance measures.

– **Criterian 3 – Top-performing classification techniques:** Recent research finds that the choice of classification techniques has a large impact on the performance of defect prediction models [39]. Thus, we select to study only the top-ranked classification techniques (see Section 7).

To satisfy criterion 1, we began our study using the 26 configurable classification techniques from Table 1. To satisfy criterion 2, we include only 9 of the 26 classification techniques that yield the largest performance improvement when automated parameter optimization is applied (see Figure 4a), i.e., C5.0, AdaBoost, AVNNet, CART, PCANNet, NNet, FDA, MLPWeightDecay, and MLP classification techniques. To satisfy criterion 3, we select only the top 10 classification techniques that are likely to appear at the top-rank classification techniques (see Figure 16). Thus, we exclude 5 classification techniques that infrequently or never appear at the top-ranked classification techniques (i.e., AdaBoost, CART, PCANNet, FDA, MLPWeightDecay, MLP). Table 1 provides an overview of the 4 classification techniques that satisfy our criteria for analysis (i.e., C5.0, xGBTree, AVNNet, and GBM), which have varying types and numbers of parameters.

In order to measure the impact that automated parameter optimization techniques have on defect prediction models, we train defect prediction models using the optimal settings that are generated by the 4 studied optimization techniques and the default settings. To ensure that the training and testing corpora have similar characteristics, we do not re-balance or re-sample the training data.

To address RQ7, for each of the 12 performance measure, we start with the performance distribution of the 4 studied classification techniques. For each classification technique, we compute the difference in the performance of classifiers that are trained using default settings and optimal parameter settings that are generated by the studied optimization techniques (i.e., grid search, random search, genetic algorithm, and differential evolution).

### 8.2.3 Results

To analyze the difference of the performance of the studied classification techniques when applying the 4 studied automated parameter optimization techniques, we use boxplots of Figure 13 to present the distribution of the performance difference ratio (i.e., the performance of the optimized classifiers that are trained with optimal settings of random search, genetic algorithm, and differential evolution) divided by the performance of the grid-search optimized classifiers) when apply to the studied datasets and the studied classification techniques.

**Grid search, random search, genetic algorithm, and differential evolution techniques yield similar benefits of AUC performance improvement when applying to defect prediction models.** Figure 13 shows the distributions of the performance difference ratio for the 12 performance measures. Irrespective of the choice of performance measures, the improvement ratios that are centerred at one indicate that optimized defect prediction models tend to produce similar performance irrespective of the used automated parameter optimization techniques. The actual AUC values of defect prediction models that are trained with default settings and optimized settings (see Figure 15) also confirm that defect prediction models that are optimized by grid search, random search, genetic algorithm, and differential evolution tend to produce similar AUC performance values. This result suggests that the choice of automated parameter optimization techniques does not pose a great threat to the conclusions of defect prediction studies. Thus, practitioners and researchers can safely adopt any of the four studied automated parameter optimization techniques for defect

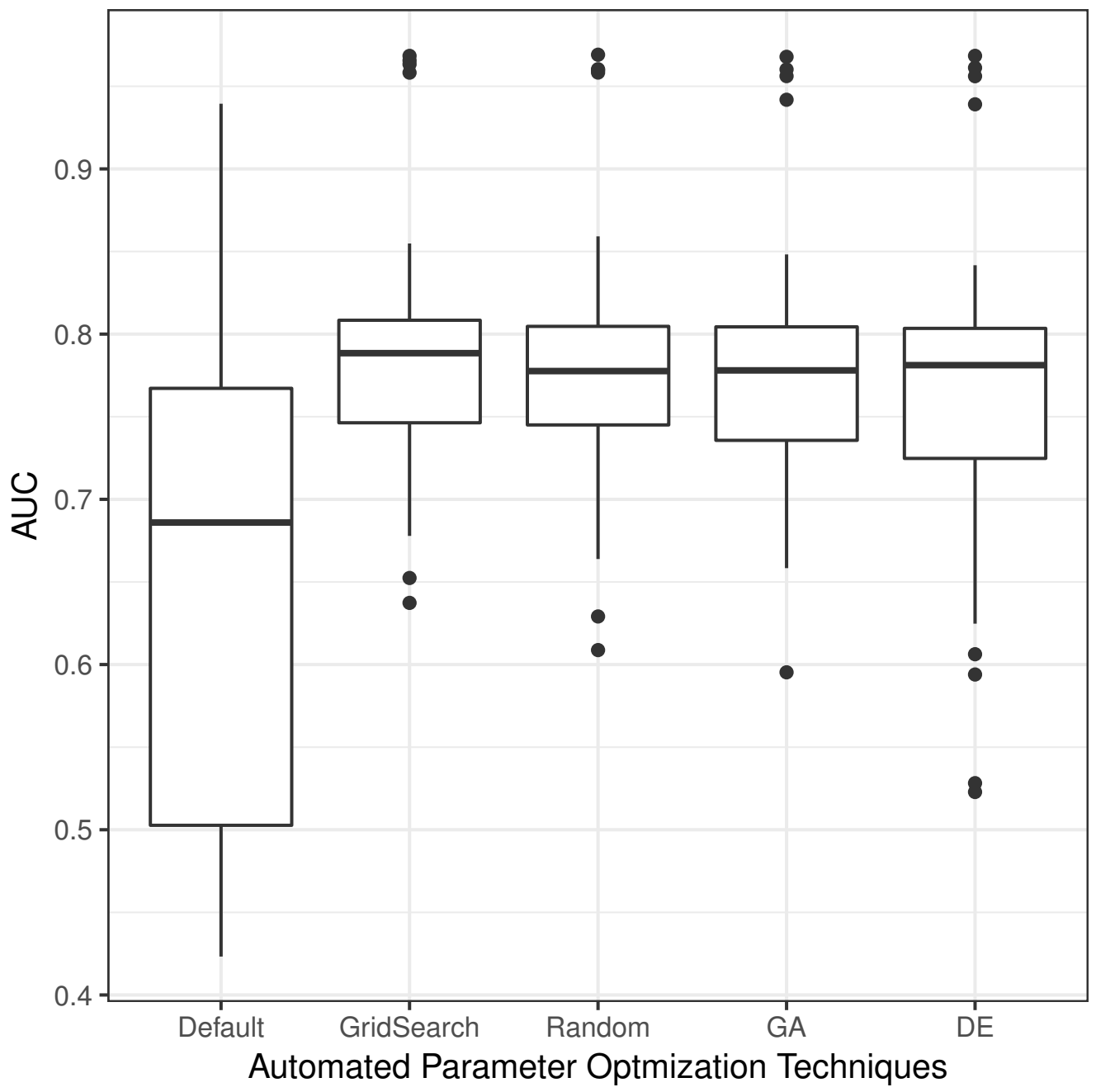

Fig. 14: The distributions of the actual AUC performance values of default-settings models and grid search (GridSearch), random search (Random), genetic algorithm (GA), and differential evolution (DE) optimized models.

prediction models. However, when other dimensions like computational cost are considered, our recommendation might be changed.

*Irrespective of the choice of performance measures, grid search, random search, genetic algorithm, and differential evolution techniques yield similar benefits of performance improvement when optimizing defect prediction models. This result suggests that, with respect to the performance improvement, practitioners and researchers can safely adopt any of the four studied automated parameter optimization techniques for defect prediction models. However, when other dimensions like computational cost are considered, our recommendation might be changed.*

# 9 DISCUSSION

In this section, we discuss the broader implications of our findings and how our findings fit with (1) the conclusions of prior defect prediction studies; (2) the experimental design of defect prediction models; and (3) search-based software engineering literature.

## 9.1 Conclusions of Prior Defect Prediction Studies

In the past 10 years, there has been a spike in the use of random forest classification technique in defect prediction studies [37, 40, 43, 45, 53, 81, 117, 138, 150]. For example, Guo *et al.* [43] and Lessmann *et al.* [81] show that a random forest classifer is the top-performing classification techniques. Fukushima *et al.* [37] developed a Just-In-Time defect prediction model using random forest classifiers.

Furthermore, the conclusions of prior defect prediction studies heavily rely on random forest classifiers. For example, recent studies use a random forest classification technique to investigate (1) the impact of issue report mislabelling on defect prediction models [138]; (2) the impact of model validation techniques on defect prediction models [141]; (3) the impact of the discretization of continuous defect counts [117]; (4) the impact of feature selection techniques on defect prediction models [40, 150]; (5) the impact of local-global defect prediction models [53]; and (6) the privacy and utility for defect prediction models [108–110].

**Implications.** The automated parameter optimization of random forest classifiers (1) has little impact on the AUC performance improvement and (2) its stability; and (3) has little impact on the interpretation of defect prediction models. This result indicates that automated parameter optimization does not pose a threat to the conclusions of prior defect prediction studies (which make use of random forest classifiers), suggesting that researchers and practitioners should not be too concerned about the impact of parameter settings on random forest classifiers.

## 9.2 The Experimental Design of Defect Prediction Models

Plenty of research raises concerns about the impact of experimental components on the conclusions of defect prediction studies [135]. For example, concerns about the quality of defect datasets have been raised [14, 103, 115, 138, 149]. Bird *et al.* [14] find that the issue reports of several defects are not mentioned in commit logs. Bachmann *et al.* [6] find that the noise generated by missing links in defect prediction datasets introduces bias. Kim *et al.* [72] find that the randomly-generated noise has a large negative impact on the performance of defect models. On the other hand, our recent work [138] shows that realistic noise (i.e., noise generated by actually mislabelled issue reports [54]) does not typically impact the precision of defect prediction models.

Recent research also raises concerns about that the choice of classification and model validation techniques. For example, Ghotra *et al.* [39] find that there are statistically significant differences in the performance of defect prediction models that are trained using different classification techniques. Panichella *et al.* [105] and Bowes *et al.* [16] also find that using different classification techniques can identify different defective modules. Our prior work [141] shows that the choice of model validation technique also has an impact on the accuracy and stability of performance estimates.

**Implications.** In addition to the concerns about data quality, classification techniques, and model validation techniques, we find that the parameter settings of classification techniques can also have a large impact on the performance, stability, and interpretation of defect prediction models. Our findings suggest that researchers should experiment with the parameters of the classification techniques. Given the large parameter space of many classification techniques, automated parameter optimization techniques like Caret offer an efficient way to conduct such an exploration. While there are many classification techniques like random forest for which the impact of automated parameter optimization is negligible or small, such optimization is still simple to adopt in practice given the availability of automated parameter optimization in commonly-used research toolkits (e.g., Caret package for R [78], GA package for R [122], DEoptim package for R [100], MultiSearch for Weka [44], GridSearch for Scikit-learn [106]).

### 9.3 Parameter Optimization for Defect Prediction Models

The most closely research related to our paper is the work by Fu *et al.* [35], who demonstrate that automated parameter optimization has an impact on the performance, model interpretation, and the ranking of top-performing classification techniques. However, their experimental design is different from our paper in numerous ways. First, we investigate the impact of parameter settings of 26 classification techniques, while they focus on only 4 classification techniques. Second, we use 12 performance measures for our evaluation, namely, 3 threshold-independent and 9 threshold-dependent performance measures (RQ1), while they focus on only 3 threshold-dependent measures (i.e., precision, recall, and f-measure). Finally, we apply four off-the-shelf parameter optimization techniques, namely, grid search, random search, genetic algorithm, and differential evolution, while they apply only differential dvolution (DE) algorithm.

**Implications.** Even though the design of our experiments differ, our conclusions (i.e., RQ1, RQ3, and Section 7) are consistent to the conclusions of their study. Relative to the contributions of prior work, this paper makes three additional contributions: an investigation of the impact of automated parameter optimization on model stability (RQ2), an investigation of parameter transferability across datasets of software systems (RQ4), and an empirical comparison of automated parameter optimization techniques for defect prediction models (Section 8.2.3).

### 9.4 Search-Based Software Engineering

Prior studies in the area of the search-based software engineering show that search algorithms can be used to find useful solution to software engineering problems [46, 47]. For example, Panichella *et al.* [104] and Lohar *et al.* [84] tune the parameters of topic modelling techniques (i.e., Latent Dirichlet Allocation) for software engineering tasks. Thomas *et al.* [143] tune off-the-shelf Information Retrieval techniques for bug localization. Song *et al.* [128] tune classification techniques for software effort estimation models. Wang *et al.* [147] tune the settings of code clone detection techniques. Arcuri *et al.* [3] investigate a minimal test suite that maximizes branch coverage in the context of test data generation. Jia *et al.* [60] investigate an optimal test suite for combinatorial interaction testing. Fu *et al.* [34] tune the parameter settings of deep learning for software engineering.

**Implications.** In addition to the area of topic modelling, information retrieval, effort estimation, code clone detection, and test data and suite generation, we find that automated parameter optimization can also be used to efficiently tune parameter settings of classification techniques for defect prediction models. Furthermore, we find that optimization yields a large benefit in terms of the performance of defect prediction models, suggesting that automated parameter optimization should be used by future defect prediction studies.

## 10 Threats to Validity

We now discuss the threats to the validity of our study.

### 10.1 Construct Validity

The datasets that we analyze are part of several corpora (e.g., NASA and PROMISE), which each provide different sets of metrics and different granularity of modules (e.g., method, class, directory). Since the metrics and granularity of modules vary, this is a point of variation between the studied systems that could impact our results. However, our analysis on datasets that have the same set of metrics shows that the number and type of predictors do not influence our findings. Thus, we conclude that the variation of metrics does not pose a threat to our study. On the other hand, the variety of metrics and granularity of modules also strengthens the generalization of our results, i.e., our observations are not bound to one specific set of metrics.

The grid-search budget, which controls the number of settings that we evaluate for each parameter, limits our exploration of the parameter space. Although our budget setting is selected based on the literature [78], selecting a different budget may yield different results. To combat this bias, we repeat the experiment with different budgets of exploration (i.e., 3, 5, 7) and find consistent conclusions. Thus, we believe that an increase in the budget would not alter the conclusions of our study.

Automated parameter optimization techniques also have configurable parameters that determine the characteristics of the optimization algorithms (e.g., the `budget` parameter of the grid search technique). Prior work in the area of search-based software engineering point out that the parameters of search-based techniques have a large impact on the performance of software engineering applications For example, Arcuri *et al.* [2, 3] show that the parameters of search-based techniques substantially change the performance of test-case data generation. Indeed, the use of different parameter settings of automated parameter optimization techniques may produce different parameter space for evaluation. However, the exploration of the parameter space of automated parameter optimization techniques on a large number of classification techniques and datasets may require a large computational cost. Thus, future research should explore the impact of the parameters of automated parameter optimization techniques on defect prediction models.

While previous research [4, 11, 15] suggested mitigating the skewness distribution of defect datasets, log transformation may not be necessary for all classification techniques (except logistic regression). Indeed, Jiang *et al.* [63] point out that log transformation rarely affects the performance of defect prediction models. Thus, we suspect that the use of log transformation may not pose a threat to the validity of our conclusions. However, applying other choices of data transformation techniques may yield different results.

### 10.2 Internal Validity

Prior work shows that noisy data may influence conclusions that are drawn from defect prediction studies [39, 135, 138]. Hence, noisy data may be influencing our conclusions. However, we conduct a highly-controlled experiment where known-to-be noisy NASA data [124] has been cleaned. Recently, Jean *et al.* [111] point out that the cleaned NASA datasets that are provided by Shepperd *et al.* [124] are still problematic. To check whether this noise is impacting our conclusions, we repeat our experiment after dropping the NASA datasets (i.e., JM1 and PC5). We find that the problematic NASA datasets do not alter the conclusions of our study.

### 10.3 External Validity

We study a limited number of systems in this paper. Thus, our results may not generalize to all software systems (e.g., open-source and commercial systems). However, the goal of this paper is not to show a result that generalizes to all datasets, but rather to show that there are datasets where parameter optimization matters. Nonetheless, additional replication studies may prove fruitful.

The conclusions of the ranking of classification techniques rely on the AUC performance measure (see Section 7). While AUC is a standard performance measure for model comparison, software teams may have different objective goals which require different performance measure. When we repeat the experiment using MCC measure, we find that the use of different performance measures does not alter the conclusions of the ranking of classification techniques. Indeed, with respect to MCC measure, we find that optimized C5.0 classifiers (1) are the top-performing classification technique; and (2) outperform random forest classifiers. In addition, we find that there many of the classification techniques in the top-3 ranks overlap between AUC and MCC measures, i.e., 6 of 8 classifiers (75%) that appear in the top-3 ranks of the AUC measure also appear in the top-3 ranks of the MCC measure. We also host these results in our online appendix.[6] Furthermore, the conclusions of the ranking of classification techniques (Section 7) rely on the grid-search parameter optimization. Since we find that the four studied parameter optimization techniques yield similar benefits of performance improvement when optimizing defect prediction models (see Section 8), we suspect that the choice of parameter optimization may not pose a threat to validity.

The conclusions of the comparison of automated parameter optimization techniques rely on the performance improvement (see Section 8). However, other dimensions that should be considered when selecting automated parameter optimization techniques in practice (e.g., computational cost, customization for other performance measures, sensitivity of the parameters of optimization techniques). Recent studies raise concerns that the computational cost of grid search technique is larger than advanced optimization techniques [10, 36], e.g., random search, genetic algorithm, and differential evolution. However, it is the strength of our paper that the experimental design is highly controlled—i.e., the candidate parameter settings of grid search technique do not vary for each dataset and bootstrap sample, while the candidate parameter settings of such advanced parameter optimization techniques often vary.

Our comparison of automated parameter optimization techniques relies on the observed performance improvements (see Section 8). However, other dimensions (e.g., computational cost, customization for other performance measures, sensitivity of the parameters of optimization techniques) should be considered when selecting automated parameter optimization techniques in practice. Recent studies raise concerns that the computational cost of the grid search technique is larger than advanced optimization techniques, e.g., random search, genetic algorithm, and differential evolution [10, 36]. However, in this paper, we sought to ensure that the experimental design is highly controlledi.e., the candidate parameter settings of grid search technique do not vary for each dataset and bootstrap samples, while the candidate parameter settings of such advanced parameter optimization techniques often vary. Thus, future research should extensively investigate the strength and weakness of automated parameter optimization techniques.

The datasets that we analyze are part of several corpora (e.g., NASA and PROMISE), which each provide different sets of metrics and different granularity of modules (e.g., method, class, directory). Since the metrics and granularity of modules vary, this is a point of variation between the studied systems that could impact our results. However, our analysis on datasets that have the same set of metrics shows that the number and type of predictors do not influence our findings.

In Section 8, we study a limited number of automated parameter optimization techniques, namely, grid search, random search, genetic algorithm, and differential evolution. Our study shows that these parameter optimization techniques yield similar AUC performance when applied to defect prediction models, however, other automated parameter optimization techniques may produce different results. Nevertheless, these techniques are off-the-shelf parameter optimization techniques that are readily-available. Thus, our observations and guidances are easily transferable to industrial practices.

## 11 Conclusions

Defect prediction models are classifiers that are trained to identify defect-prone software modules. The characteristics of the classifiers that are produced are controlled by configurable parameters. Recent studies point out that classifiers may under-perform because they were trained using suboptimal default parameter settings. However, it is impractical to explore all of the possible settings in the parameter space of a classification technique.

In this paper, we investigate the (1) performance improvement, (2) performance stability, (3) model interpretation, (4) parameter transferability, (5) computational cost, and (6) the ranking of classification techniques of defect prediction models where automated parameter optimization technique (i.e., grid search) has been applied. Through

6. https://github.com/SAILResearch/appendix-parameter_optimization_extension/

a case study of 18 datasets from systems that span both proprietary and open source domains, we make the following observations:

- Optimization improves the AUC performance of defect prediction models by up to 40 percentage points. Fortunately, random forest that is popularly-used in defect prediction studies tends to have negligible to small impact on the AUC performance of defect prediction models.
- Optimized classifiers are at least as stable as classifiers that are trained using the default settings. On the other hand, for random forests classifiers, the performance of the default-settings models is as stable as the optimized models.
- Optimization substantially shifts the importance ranking of software metrics, with as few as 28% of the top-ranked variables from the optimized models appearing in those ranks in the default-settings models. On the other hand, 95% of the variables appear at the same rank in the classification techniques where optimization has a negligible performance improvement (e.g., random forest).
- The optimized settings of 17 of the 20 most sensitive parameters can be applied to datasets that share a similar set of metrics without a statistically significant drop in performance. On the other hand, we find that only the parameters of LogitBoost, FDA, and xGBTree cannot be transferred across any of the studied datasets.
- Grid-search optimization adds less than 30 minutes of additional computation time to 12 of the 26 studied classification techniques.
- Random forest is not the top-performing classification technique for all of the studied datasets, which disagrees with the findings of data mining studies. In addition, some rarely-used low-performing classification techniques (e.g., C5.0) are substantially outperforming widely-used techniques like random forest — highlighting the importance of exploring the parameter settings of classifiers.

Then, we investigate which automated parameter optimization techniques yield the largest performance improvement for defect prediction models. Based on an empirical comparison of the four state-of-the-art automated parameter optimization techniques, namely, grid search, random search, genetic algorithm, and differential evolution, we find that:

- Irrespective of the choice of performance measures, automated parameter optimization techniques like grid search, random search, genetic algorithm, and differential evolution yield similar benefits of performance improvement when applying to defect prediction models.

Our results lead us to conclude that automated parameter optimization can have a large impact on the performance improvement, performance stability, model interpretation, and ranking of defect prediction models. However, there are some classification techniques like random forest and support vector machines that are not impacted by such optimization. Thus, future studies should apply automated parameter optimization only if the used classification techniques are sensitive to parameter settings, and we should not be too concerned about the most commonly-used random forests. Since we find that random forest tends to be the top-performing classification techniques, tends to produce stable performance estimates, and is insensitive to parameter settings, future studies should consider using a random forest classifier when constructing defect prediction models.

Finally, we would like to emphasize that we do not seek to claim the generalization of our results. Instead, the key message of our study is to shed light that automated parameter optimization should be considered in many classification techniques like a neural network, but it is not essential in some classification techniques like a random forest. Nonetheless, some rarely-used low-performing classification techniques (e.g., C5.0) are substantially outperforming widely-used techniques like random forest when the parameters of C5.0 classifiers are optimized—highlighting the importance of exploring the parameter settings of classifiers. Hence, we recommend that software engineering researchers experiment with automated parameter optimization instead of relying on default parameter setting of research toolkits. Given the availability of automated parameter optimization in commonly-used research toolkits (e.g., Caret package for R [78], GA package for R [122], DEoptim package for R [100], MultiSearch for Weka [44], GridSearch for Scikit-learn [106]), we believe that our recommendation is a rather simple and low-cost recommendation to adopt.

### 11.1 Future Work

In this section, we outline research avenues for future work.

- **Investigating the impact of automated parameter optimization techniques on the ranking of classification techniques.** The conclusions of the ranking of classification techniques rely on grid search parameter optimization (see Section 7). Since we find that the four studied parameter optimization techniques yield similar performance improvements when optimizing defect prediction models (see Section 8), we suspect that the choice of parameter optimization may not pose a threat to validity. Nevertheless, future research should investigate the impact of automated parameter optimization techniques on the ranking of classification techniques.
- **Investigating the impact of automated parameter optimization techniques on ensemble learning.** Ensemble learning is a classifier that combines predictions from multiple algorithms (e.g., boosting, bucketing, and stacking). Such advanced classifiers also have configurable parameters. Thus, future study should explore if automated parameter optimization techniques improve the overall performance of such ensemble classifiers.

## Acknowledgments

This study would not have been possible without the data shared in the Tera-PROMISE repository [96] (curated by Tim

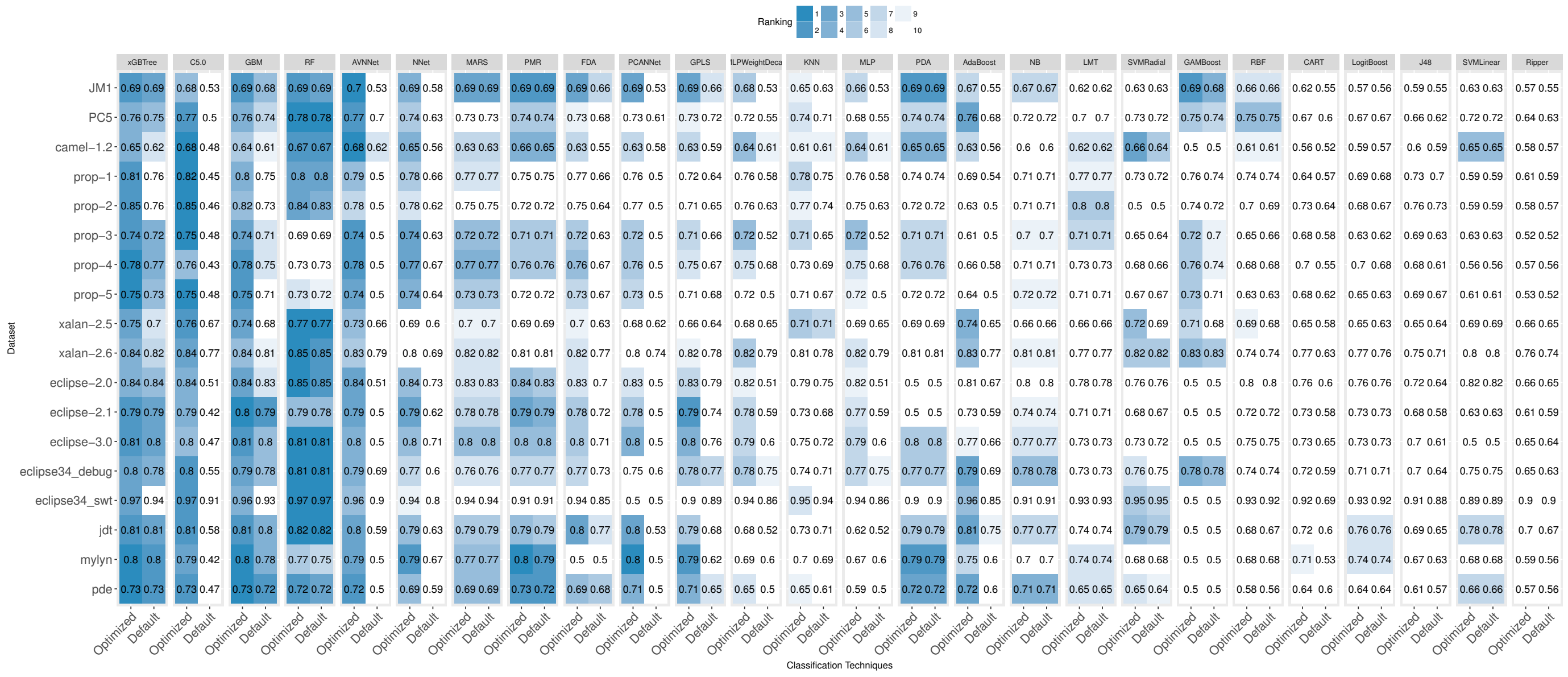

Fig. 15: The AUC performance of optimized models vs default-setting models for each classification technique.

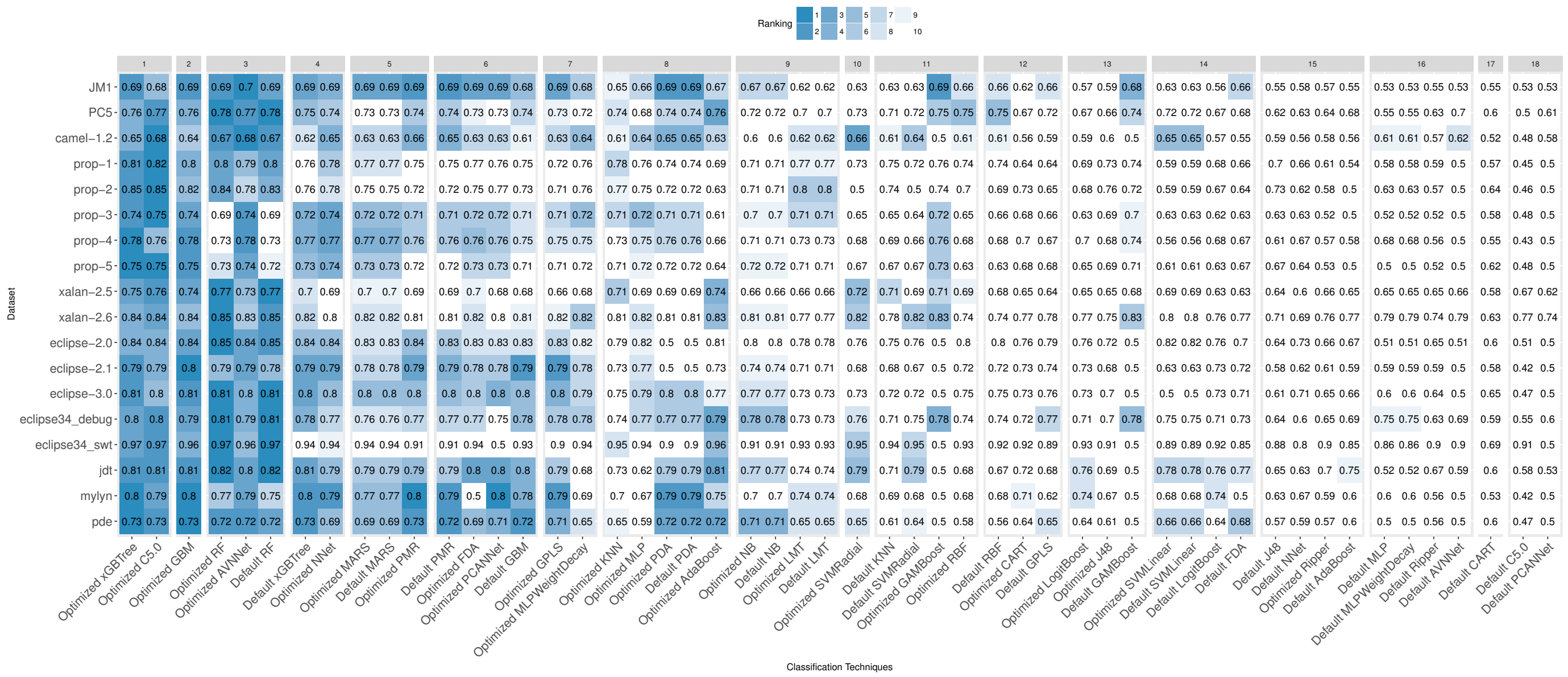

Fig. 16: The AUC performance of the ranking of the optimized models vs default-setting models

Menzies, Mitch Rees-Jones, Rahul Krishna and Carter Pape), as well as the data shared by Shepperd *et al.* [124], Zimmermann *et al.* [155], Kim *et al.* [72, 149], and D'Ambros *et al.* [23, 24], as well as High Performance Computing (HPC) systems that are provided by the Compute Canada[7] and the Centre for Advanced Computing at Queen's University.[8] This work was supported by the JSPS Program for Advancing Strategic International Networks to Accelerate the Circulation of Talented Researchers: Interdisciplinary Global Networks for Accelerating Theory and Practice in Software Ecosystems, the Grant-in-Aid for JSPS Fellows (No. 16J03360), and the Natural Sciences and Engineering Research Council of Canada (NSERC).

7. http://www.computecanada.ca
8. http://cac.queensu.ca/

## Appendix

Figure 15 shows the AUC performance of optimized models vs default-setting models for each classification technique. Figure 16 shows the AUC performance of the ranking of the optimized models vs default-setting models

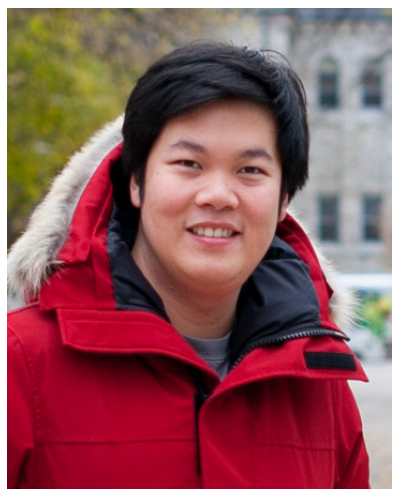

**Chakkrit Tantithamthavorn** is a lecturer at the School of Computer Science, the University of Adelaide, Australia. Prior to that, he was a research fellow at Queen's University and Nara Institute of Science and Technology. During his Ph.D. study, he won one of the most prestigious and selective sources of national funding in Japan, i.e., a JSPS Research Fellowship for Young Researchers and a Grants-in-Aid for JSPS Fellow, and won the "Best Ph.D. Student Award". His work has been published at several top-tier software engineering venues, such as the IEEE Transactions on Software Engineering (TSE), the Springer Journal of Empirical Software Engineering (EMSE) and the International Conference on Software Engineering (ICSE). His Ph.D. thesis aims to improve the fundamentals of analytical modelling for software engineering in order to produce more accurate predictions and reliable insights. His research interests include empirical software engineering and mining software repositories (MSR). He received the B.E. degree in computer engineering from Kasetsart University, Thailand, the M.E. and Ph.D. degrees in Information Science from Nara Institute of Science and Technology, Japan. More about Chakkrit and his work is available online at http://chakkrit.com.

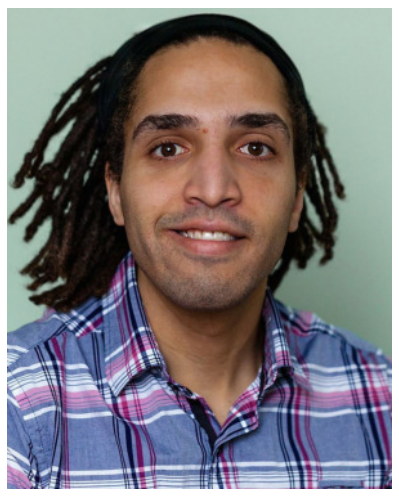

**Shane McIntosh** is an assistant professor in the Department of Electrical and Computer Engineering at McGill University. He received his Bachelor's degree in Applied Computing from the University of Guelph and his MSc and PhD in Computer Science from Queen's University. In his research, Shane uses empirical software engineering techniques to study software build systems, release engineering, and software quality. His research has been published at several top-tier software engineering venues, such as the International Conference on Software Engineering (ICSE), the International Symposium on the Foundations of Software Engineering (FSE), and the Springer Journal of Empirical Software Engineering (EMSE). More about Shane and his work is available online at http://shanemcintosh.org/.

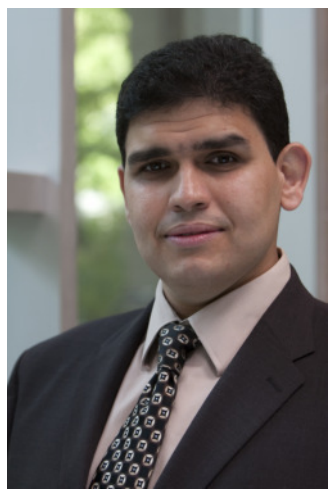

**Ahmed E. Hassan** is the Canada Research Chair (CRC) in Software Analytics, and the NSERC/BlackBerry Software Engineering Chair at the School of Computing at Queens University, Canada. His research interests include mining software repositories, empirical software engineering, load testing, and log mining. He received a PhD in Computer Science from the University of Waterloo. He spearheaded the creation of the Mining Software Repositories (MSR) conference and its research community. He also serves on the editorial boards of IEEE Transactions on Software Engineering, Springer Journal of Empirical Software Engineering, and PeerJ Computer Science. More about Ahmed and his work is available online at http://sail.cs.queensu.ca/.

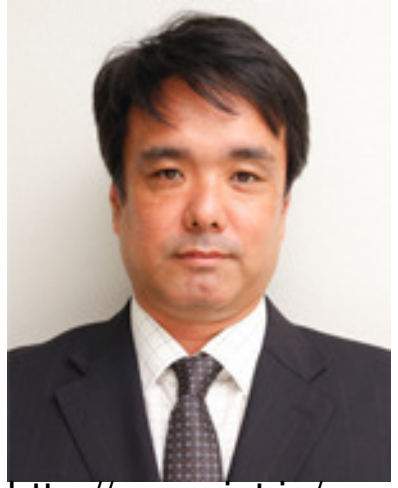

**Kenichi Matsumoto** is a professor in the Graduate School of Information Science at Nara Institute of Science and Technology, Japan. He received the Ph.D. degree in information and computer sciences from Osaka University. His research interests include software measurement and software process. He is a fellow of the IEICE, a senior member of the IEEE, and a member of the ACM, and the IPSJ. More about Kenichi and his work is available online at http://se-naist.jp/.